\documentclass[a4paper,11pt, english]{article} 
\usepackage{jinstpub} 
\usepackage{babel}
\usepackage{graphbox}
\usepackage{multirow}
\usepackage{makecell}
\usepackage{amsmath,amssymb, gensymb}
\usepackage{natbib,hyperref}
\usepackage{url}
\usepackage{microtype}
\usepackage{subfig}
\usepackage[useregional]{datetime2}

\usepackage{siunitx}
\usepackage{xspace}

\newcommand{\NEW}{NEXT-White}

\newcommand{\fig}{figure}
\newcommand{\eq}{equation}
\newcommand{\tab}{table}

\newcommand{\ie}{{\it i.e.}}

\newcommand{\micro}{\ensuremath{\mu}}

\newcommand{\stat}{\textrm{(stat.)}}
\newcommand{\sys}{\textrm{(sys.)}}

\newcommand{\bbonu}{\ensuremath{\beta\beta0\nu}}

\newcommand{\tz}{\ensuremath{t_0}}
\newcommand{\st}{\ensuremath{S_2}}
\newcommand{\so}{\ensuremath{S_1}}

\newcommand{\XE}{\ensuremath{{}^{136}\rm Xe}}

\newcommand{\Kr}[1]{\ensuremath{^{#1}\mathrm{Kr}}\xspace}

\newcommand{\Rb}[1]{\ensuremath{^{#1}\mathrm{Rb}}\xspace}

\DeclareSIUnit\c{\mbox{$c$}}
\DeclareSIUnit\magn{\mbox{$\times$}}
\DeclareSIUnit\min{min}
\DeclareSIUnit\week{week}
\DeclareSIUnit\year{yr}
\DeclareSIUnit\years{years}
\DeclareSIUnit\yr{yr}
\DeclareSIUnit\standard{std}
\DeclareSIUnit\str{sr}
\DeclareSIUnit\ppm{ppm}
\DeclareSIUnit\ppb{ppb}
\DeclareSIUnit\ppt{ppt}
\DeclareSIUnit\pe{PE}
\DeclareSIUnit\spe{SPE}
\DeclareSIUnit\ev{events}
\DeclareSIUnit\ct{counts}
\DeclareSIUnit\neutron{\mbox{$n$}}
\DeclareSIUnit\smp{samples}
\DeclareSIUnit\Sample{S}
\DeclareSIUnit\ch{ch}
\DeclareSIUnit\hit{hit}
\DeclareSIUnit\hits{hits}
\DeclareSIUnit\bin{(\mbox{5-PE}~bin)}
\DeclareSIUnit\sgm{\mbox{$\sigma$}}
\DeclareSIUnit\rms{RMS}
\DeclareSIUnit\keVr{\mbox{keV$_{\rm nr}$}}
\DeclareSIUnit\keVee{\mbox{keV$_{e{\rm e}}$}}
\DeclareSIUnit\ph{photon}
\DeclareSIUnit\pes{pes}
\DeclareSIUnit\el{electrons}
\DeclareSIUnit\pm{PMT}
\DeclareSIUnit\inch{"}
\DeclareSIUnit\bit{bit}
\DeclareSIUnit\sample{samples}
\DeclareSIUnit\barn{barn}
\DeclareSIUnit\bara{bar}
\DeclareSIUnit\barg{barg}
\DeclareSIUnit\mlardepth{\mbox(meter~of~\LAr~depth)}
\DeclareSIUnit\Curie{Ci}
\DeclareSIUnit\psi{psi}
\DeclareSIUnit\parsec{pc}
\DeclareSIUnit\liveday{\mbox{live-days}}
\DeclareSIUnit\days{\mbox{days}}
\DeclareSIUnit\day{\mbox{day}}
\DeclareSIUnit\miles{\mbox{miles}}
\DeclareSIUnit\degreeC{\mbox{$^{\circ}$C}}
\DeclareSIUnit\electron{\mbox{$e^-$}}
\DeclareSIUnit\Euro{\mbox{\euro}}
\DeclareSIUnit\cph{cph}
\DeclareSIUnit\neq{neq}
\DeclareSIUnit\unit{unit}
\DeclareSIUnit\byte{Byte}
\DeclareSIUnit\Bq{\becquerel}

\newcommand{\HPXeEL}{HPXe-EL}

\newcommand{\RII}{Run II}

\newcommand{\XenonQbb}{\SI{2458}{\keV}}

\newcommand{\XenonLongitudinalDiffusion}{\ensuremath{0.3\ \textrm{mm}/\sqrt{\textrm{cm}}}}
\newcommand{\XenonTransverseDiffusion}{\ensuremath{1\ \textrm{mm}/\sqrt{\textrm{cm}}}}

\newcommand{\simto}{\ensuremath{\sim}}

\newcommand{\VDR}{\ensuremath{v_d}}
\newcommand{\LT}{\ensuremath{\tau}}

\newcommand{\E}{\ensuremath{E}}
\newcommand{\PR}{\ensuremath{P}}
\newcommand{\TPT}{\ensuremath{T}}
\newcommand{\TOX}{\ensuremath{T_0}}

\newcommand{\ZP}{\ensuremath{Z^P}}
\newcommand{\ZPOX}{\ensuremath{Z^P_0}}

\newcommand{\Z}{\ensuremath{z}}

\newcommand{\XY}{\ensuremath{(x, y)}}

\newcommand{\NewTo}{\SI{293.15}{\kelvin}}
\newcommand{\NewSevenBarPz}{\num{0.963}}
\newcommand{\NewNineBarPz}{\num{0.953}}
\newcommand{\NewPo}{\num{0.995}}

\newcommand{\StandardLDRunII}{\ensuremath{318.9 \pm 1.8\ \stat\ \pm 20.1\ \sys\ \micro \textrm{m}/\sqrt{\textrm{cm}}}}
\newcommand{\StandardTDRunII}{\ensuremath{1279 \pm 3\ \stat\ \pm 40\ \sys\ \micro \textrm{m}/\sqrt{\textrm{cm}}}}
\newcommand{\LDTenBarRunII}{\ensuremath{267.3 \pm 1.5\ \stat\ \pm 16.9\ \sys\ \micro \textrm{m}/\sqrt{\textrm{cm}}}}
\newcommand{\TDTenBarRunII}{\ensuremath{1072 \pm 3\ \stat\ \pm 34\ \sys\ \micro \textrm{m}/\sqrt{\textrm{cm}}}}

\newcommand{\MCDriftVelocity}{\SI{1}{\mm\per\micro\second}}
\newcommand{\MCDriftVelocityFitRunII}{\SI[parse-numbers=false]{999.79 \pm 0.05\ \stat\ \pm 1.88\ \sys}{\micro\meter\per\micro\second}}
\newcommand{\DriftVelocityStandardRunII}{\SI[parse-numbers=false]{967.99 \pm 0.17\ \stat\ \pm 4.06\ \sys}{\micro\meter\per\micro\second}}

\newcommand{\ELVelocitySevenBarRunII}{\SI{3.72 +- 0.03}{\mm\per\micro\second}}
\newcommand{\ELVelocityNineBarRunII}{\SI{3.52 +- 0.03}{\mm\per\micro\second}}

\newcommand{\TimeToCrossELSevenBarRunII}{\SI{0.806 +- 0.006}{\micro\second}}
\newcommand{\TimeToCrossELNineBarRunII}{\SI{0.852 +- 0.008}{\micro\second}}

\newcommand{\MaxDriftTimeMC}{\SI{533.12 +- 0.026}{\micro\second}}
\newcommand{\MaxDriftTimeStandardRunII}{\SI{548.64 +- 0.1}{\micro\second}}

\newcommand{\KrEnergy}{\SI{41.5}{\keV}}
\newcommand{\KrEnergyI}{\SI{32.1}{\keV}}
\newcommand{\KrEnergyII}{\SI{9.4}{\keV}}
\newcommand{\RbLifetime}{\SI{86.2}{\days}}
\newcommand{\RbIntensity}{\SI{1}{\kilo\becquerel}}
\newcommand{\KrLifetime}{\SI{1.83}{\hour}}
\newcommand{\KrLifetimeShort}{\SI{154.4}{\nano\second}}
\newcommand{\KrRate}{\SI{10}{\hertz}}

\newcommand{\KrSiPMEminRunII}{\SI{5}{\pes}}

\newcommand{\KrSearchWindow}{\SI{620}{\micro\second}}

\newcommand{\KrFidVolumeRRunII}{\SI{150}{\mm}}

\newcommand{\AmbientNormalTemperature}{\SI{20}{\celsius}}
\newcommand{\StandardNormalTemperature}{\SI{293.15}{\kelvin}}

\newcommand{\KryptonSevenBarStartRunII}{\DTMdisplaydate{2017}{9}{13}{-1}}
\newcommand{\KryptonSevenBarEndRunII}{\DTMdisplaydate{2017}{9}{19}{-1}}
\newcommand{\KryptonNineBarStartRunII}{\DTMdisplaydate{2017}{10}{22}{-1}}
\newcommand{\KryptonNineBarEndRunII}{\DTMdisplaydate{2017}{10}{29}{-1}}

\newcommand{\NewGateVoltageSevenBarRunII}{\SI{7.0}{\kV}}
\newcommand{\NewGateVoltageNineBarRunII}{\SI{8.5}{\kV}}

\newcommand{\NewCathodeVoltageSevenBarRunII}{\SI{28}{\kV}}
\newcommand{\NewCathodeVoltageNineBarRunII}{\SI{30}{\kV}}

\newcommand{\NewCathodeVoltageKrDiffSevenBarMinRunII}{\SI{16}{\kV}}
\newcommand{\NewCathodeVoltageKrDiffSevenBarMaxRunII}{\SI{28}{\kV}}

\newcommand{\NewCathodeVoltageKrDiffNineBarMinRunII}{\SI{21.5}{\kV}}
\newcommand{\NewCathodeVoltageKrDiffNineBarMaxRunII}{\SI{29.5}{\kV}}

\newcommand{\NewCathodeVoltageAtFifteenBar}{\SI{41}{\kV}}

\newcommand{\NewSevenBarPressureRunII}{\SI{7.2}{\bar}}
\newcommand{\NewNineBarPressureRunII}{\SI{9.1}{\bar}}

\newcommand{\NewDriftFieldRunII}{\SI{53.6}{\V\per\cm\per\bar}}

\newcommand{\NewPressureVesselMaterial}{316Ti}

\newcommand{\NewTpcLength}{\SI{664.5}{\mm}}

\newcommand{\NewTpcDriftLength}{\SI{530.3 +- 2}{\mm}}

\newcommand{\NewDriftField}{\SI{400}{\V\per\cm}}

\newcommand{\NewTpcELGap}{\SI{6}{\mm}}

\newcommand{\NewReducedField}{\SI{2.2}{\kV\per\cm\per\bar}}
\newcommand{\NewReducedFieldRunII}{\SI{1.7}{\kV\per\cm\per\bar}}

\newcommand{\NewGateVoltageAtFifteenBar}{\SI{16.2}{\kV}}

\newcommand{\NewNumberOfSiPM}{\num{1792}}

\newcommand{\NewSipmPitch}{\SI{10}{\mm}}

\newcommand{\NewNumberOfPMT}{\num{12}}

\newcommand{\NewCathodeToPMTs}{\SI{130}{\mm}}

\newcommand{\NewTpcDiameter}{\SI{454}{\mm}}
\newcommand{\NewFiducialMass}{\SI{5}{\kg}}
\newcommand{\NewFiducialMassSevenBar}{\SI{2.3}{\kg}}
\newcommand{\NewFiducialMassNineBar}{\SI{3}{\kg}}
\newcommand{\NewPressure}{\SI{15}{\bar}}
\newcommand{\NewMinimumPressure}{\SI{10}{\bar}}

\newcommand{\NewTypePMT}{Hamamatsu R11410-10}
\newcommand{\NewPMTCoverage}{31\%}

\newcommand{\NewBarrelICS}{\SI{60}{\mm}}

\newcommand{\NewPlatesICS}{\SI{120}{\mm}}

\begin{document}
\title{Electron drift properties in high pressure gaseous xenon}
\collaboration{The NEXT Collaboration}
\author[17,6,a]{A.~Sim\'on,\note[a]{Corresponding author.}}
\author[17]{R.~Felkai,}
\author[17,18]{G.~Mart\'inez-Lema,}
\author[3,14]{F.~Monrabal,}
\author[18]{D.~Gonz\'alez-D\'iaz,}
\author[17]{M.~Sorel,}
\author[18]{J.A.~Hernando~Morata,}
\author[14, 8, 17,b]{J.J.~G\'omez-Cadenas,\note[b]{NEXT Co-spokesperson.}}
\author[10]{C.~Adams,}
\author[17]{V.~\'Alvarez,}
\author[6]{L.~Arazi,}
\author[4]{C.D.R~Azevedo,}
\author[2]{K.~Bailey,}
\author[17]{J.M.~Benlloch-Rodr\'{i}guez,}
\author[12]{F.I.G.M.~Borges,}
\author[17]{A.~Botas,}
\author[17]{S.~C\'arcel,}
\author[17]{J.V.~Carri\'on,}
\author[20]{S.~Cebri\'an,}
\author[12]{C.A.N.~Conde,}
\author[17]{J.~D\'iaz,}
\author[5]{M.~Diesburg,}
\author[12]{J.~Escada,}
\author[19]{R.~Esteve,}
\author[11]{L.M.P.~Fernandes,}
\author[14, 8, 17]{P.~Ferrario,}
\author[4]{A.L.~Ferreira,}
\author[11]{E.D.C.~Freitas,}
\author[14]{J.~Generowicz,}
\author[7]{A.~Goldschmidt,}
\author[10]{R.~Guenette,}
\author[9]{R.M.~Guti\'errez,}
\author[2]{K.~Hafidi,}
\author[1]{J.~Hauptman,}
\author[11]{C.A.O.~Henriques,}
\author[9]{A.I.~Hernandez,}
\author[19]{V.~Herrero,}
\author[2]{S.~Johnston,}
\author[3]{B.J.P.~Jones,}
\author[17]{M.~Kekic,}
\author[16]{L.~Labarga,}
\author[17]{A.~Laing,}
\author[5]{P.~Lebrun,}
\author[17]{N.~L\'opez-March,}
\author[9]{M.~Losada,}
\author[10]{J.~Mart\'in-Albo,}
\author[17]{A.~Mart\'inez,}
\author[3]{A.~McDonald,}
\author[11]{C.M.B.~Monteiro,}
\author[19]{F.J.~Mora,}
\author[17]{J.~Mu\~noz Vidal,}
\author[17]{M.~Musti,}
\author[17]{M.~Nebot-Guinot,}
\author[17]{P.~Novella,}
\author[3,c]{D.R.~Nygren,\note[c]{NEXT Co-spokesperson.}}
\author[17]{B.~Palmeiro,}
\author[5]{A.~Para,}
\author[17,d]{J.~P\'erez,\note[d]{Now at Laboratorio Subterr\'aneo de Canfranc, Spain.}}
\author[17]{M.~Querol,}
\author[17]{J.~Renner,}
\author[2]{J.~Repond,}
\author[2]{S.~Riordan,}
\author[15]{L.~Ripoll,}
\author[17]{J.~Rodr\'iguez,}
\author[3]{L.~Rogers,}
\author[17]{C.~Romo Luque,}
\author[12]{F.P.~Santos,}
\author[11]{J.M.F. dos~Santos,}
\author[13,e]{C.~Sofka,\note[e]{Now at University of Texas at Austin, USA.}}
\author[13]{T.~Stiegler,}
\author[19]{J.F.~Toledo,}
\author[17]{J.~Torrent,}
\author[4]{J.F.C.A.~Veloso,}
\author[13]{R.~Webb,}
\author[13,f]{J.T.~White,\note[f]{Deceased.}}
\author[17]{N.~Yahlali}
\emailAdd{ander@post.bgu.ac.il}
\affiliation[1]{
Department of Physics and Astronomy, Iowa State University, 12 Physics Hall, Ames, IA 50011-3160, USA}
\affiliation[2]{
Argonne National Laboratory, Argonne, IL 60439, USA}
\affiliation[3]{
Department of Physics, University of Texas at Arlington, Arlington, TX 76019, USA}
\affiliation[4]{
Institute of Nanostructures, Nanomodelling and Nanofabrication (i3N), Universidade de Aveiro, Campus de Santiago, Aveiro, 3810-193, Portugal}
\affiliation[5]{
Fermi National Accelerator Laboratory, Batavia, IL 60510, USA}
\affiliation[6]{
Nuclear Engineering Unit, Faculty of Engineering Sciences, Ben-Gurion University of the Negev, P.O.B. 653, Beer-Sheva, 8410501, Israel}
\affiliation[7]{
Lawrence Berkeley National Laboratory (LBNL), 1 Cyclotron Road, Berkeley, CA 94720, USA}
\affiliation[8]{
IKERBASQUE, Basque Foundation for Science, Bilbao, E-48013, Spain}
\affiliation[9]{
Centro de Investigaci\'on en Ciencias B\'asicas y Aplicadas, Universidad Antonio Nari\~no, Sede Circunvalar, Carretera 3 Este No.\ 47 A-15, Bogot\'a, Colombia}
\affiliation[10]{
Department of Physics, Harvard University, Cambridge, MA 02138, USA}
\affiliation[11]{
LIBPhys, Physics Department, University of Coimbra, Rua Larga, Coimbra, 3004-516, Portugal}
\affiliation[12]{
LIP, Department of Physics, University of Coimbra, Coimbra, 3004-516, Portugal}
\affiliation[13]{
Department of Physics and Astronomy, Texas A\&M University, College Station, TX 77843-4242, USA}
\affiliation[14]{
Donostia International Physics Center (DIPC), Paseo Manuel Lardizabal, 4, Donostia-San Sebastian, E-20018, Spain}
\affiliation[15]{
Escola Polit\`ecnica Superior, Universitat de Girona, Av.~Montilivi, s/n, Girona, E-17071, Spain}
\affiliation[16]{
Departamento de F\'isica Te\'orica, Universidad Aut\'onoma de Madrid, Campus de Cantoblanco, Madrid, E-28049, Spain}
\affiliation[17]{
Instituto de F\'isica Corpuscular (IFIC), CSIC \& Universitat de Val\`encia, Calle Catedr\'atico Jos\'e Beltr\'an, 2, Paterna, E-46980, Spain}
\affiliation[18]{
Instituto Gallego de F\'isica de Altas Energ\'ias, Univ.\ de Santiago de Compostela, Campus sur, R\'ua Xos\'e Mar\'ia Su\'arez N\'u\~nez, s/n, Santiago de Compostela, E-15782, Spain}
\affiliation[19]{
Instituto de Instrumentaci\'on para Imagen Molecular (I3M), Centro Mixto CSIC - Universitat Polit\`ecnica de Val\`encia, Camino de Vera s/n, Valencia, E-46022, Spain}
\affiliation[20]{
Laboratorio de F\'isica Nuclear y Astropart\'iculas, Universidad de Zaragoza, Calle Pedro Cerbuna, 12, Zaragoza, E-50009, Spain}

\abstract{Gaseous time projection chambers (TPC) are a very attractive detector technology for particle tracking. Characterization of both drift velocity and diffusion is of great importance to correctly assess their tracking capabilities. 

NEXT-White is a High Pressure Xenon gas TPC with electroluminescent amplification, a 1:2 scale model of the future NEXT-100 detector, which will be dedicated to neutrinoless double beta decay searches. \NEW\ has been operating at Canfranc Underground Laboratory (LSC) since December 2016. The drift parameters have been measured using \Kr{83m} for a range of reduced drift fields at two different pressure regimes, namely \NewSevenBarPressureRunII\ and \NewNineBarPressureRunII. The results have been compared with \textit{Magboltz} simulations. Agreement at the 5\% level or better has been found for drift velocity, longitudinal diffusion and transverse diffusion.}



\arxivnumber{1804.01680} 
\collaboration{\includegraphics[height=9mm]{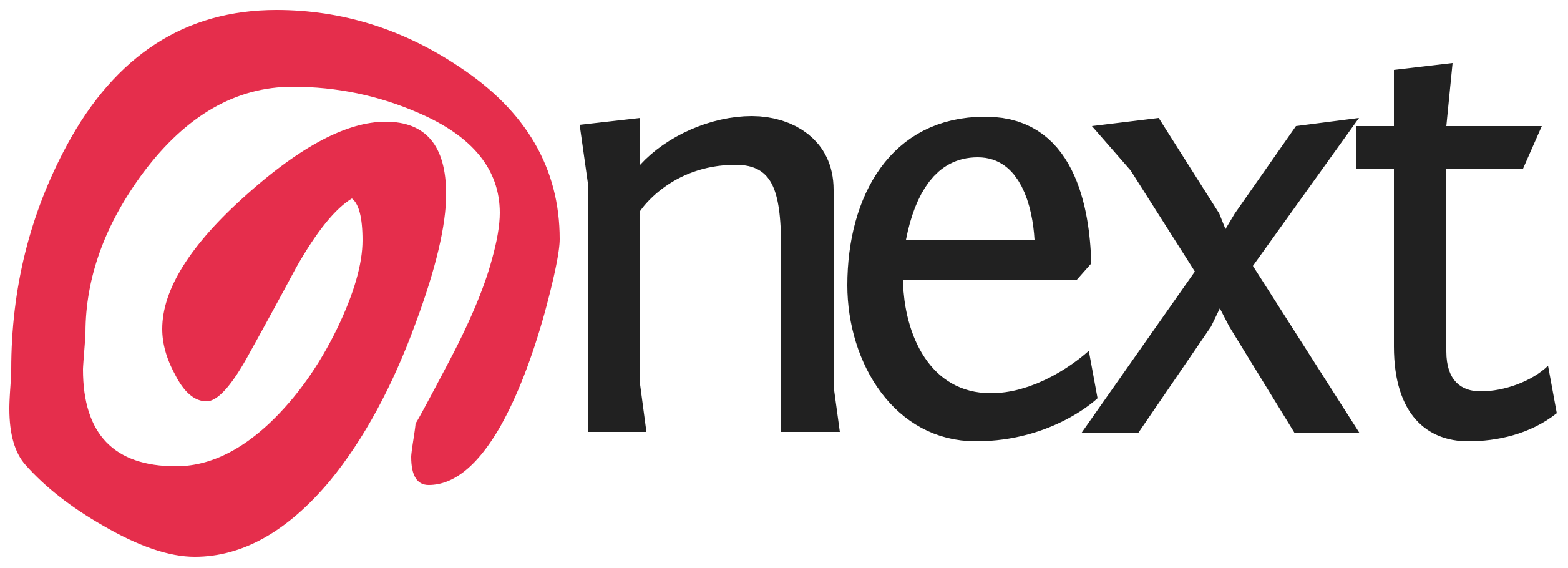}\\[6pt]
The NEXT Collaboration}

\maketitle
\flushbottom

\section{Introduction}

Electron drift properties, particularly drift velocity and diffusion, play a major role in the TPC tracking performance. In particular, the measurement of electron drift properties is of special relevance for the \NEW\ detector and its immediate successor, the NEXT-100 apparatus. Both are part of the NEXT program, which is developing the technology of high-pressure xenon  gas TPCs with electroluminescent amplification (\HPXeEL) for neutrinoless double beta decay searches \cite{Nygren:2009zz, Alvarez:2011my, Alvarez:2012haa, Gomez-Cadenas:2013lta, Martin-Albo:2015rhw}. 
The first phase of NEXT exploited two prototypes, DEMO and 
DBDM, which demonstrated the robustness of the technology, its excellent energy resolution and its particle-tracking capabilities \cite{Alvarez:2012xda, Alvarez:2013gxa, Alvarez:2012hh, Ferrario:2015kta}. The NEXT-White\footnote{Named after Prof.~James White, our late mentor and friend.} detector implements the second phase of the program. \NEW\ has been running successfully since December 2016 at the Canfranc Underground Laboratory (LSC). Its purpose is to validate the \HPXeEL\ technology in a large-scale radiopure detector. \NEW\ is a $\sim$1:2 model of NEXT-100, a 100 kg \HPXeEL\ detector, which is foreseen to start operations in 2019. 

The \NEW\ detector started operations at the LSC late in 2016. After a short engineering run (Run I) in November-December 2016, the detector was operated continuously between March and December 2017 (Run II). This paper presents the measurement of  drift parameters (drift velocity, longitudinal diffusion and transverse diffusion) for xenon at high pressure obtained during Run II. The measurements  are inferred from calibration data obtained using a rubidium source (\Rb{83}) which provides a copious sample of krypton (\Kr{83m}) decays. 

Electron transport properties in xenon gas have mostly been measured at low pressures \cite{Pack:1962, Pack:1992, Koizumi:1986, Bowe:1960, Patrick:1991, English:1953}, usually not higher than atmospheric pressure with few exceptions at high pressure \cite{Hunter:1988, Kobayashi:2006}. In this regard, the NEXT Collaboration previously measured the drift velocity and the longitudinal diffusion at 10 bar in several prototypes \cite{Alvarez:2012hh, Alvarez:2012hu, Lorca:2014sra}. However, transverse diffusion was not determined in the NEXT prototypes. In addition, measurements of the diffusion coefficients have recently been obtained by Kusano {\it et al.} \cite{Kusano:2013} at pressures ranging from \SIrange{1.7}{50}{bar}
and for reduced fields from \SIrange{6}{45}{\volt\per\cm\per\bar}. Despite 10 bar being rather far from the critical point of xenon at around 300 K (58 bar), those authors observed a strong deviation from the values expected from density-scalings, that remains unconfirmed. In order to reassess the situation, we provide here a simultaneous measurement of the three main transport observables in the range \SIrange{20}{60}{\volt\per\cm\per\bar} for pressures of \NewSevenBarPressureRunII\ and \NewNineBarPressureRunII.

The paper is organized as follows: section \ref{sec.hpxe} presents the principle of operation of a \HPXeEL\ TPC and the effect of diffusion in track reconstruction, the main features of the \NEW\ detector are summarized in 
section \ref{sec.new}, the experimental setup is described in \ref{sec.krdata}, the data selection in section \ref{sec:sel}, the measurement of the drift velocity is explained in 
section \ref{sec.dv}, the measurement of both longitudinal and transverse diffusion in section \ref{sec.diff} and, finally, conclusions are presented in section \ref{sec.conclu}

\section{Principle of operation and the effect of diffusion in \HPXeEL\ TPCs}
\label{sec.hpxe}

\begin{figure}[!htb]
  \begin{center}
    \includegraphics[width=0.45\textwidth]{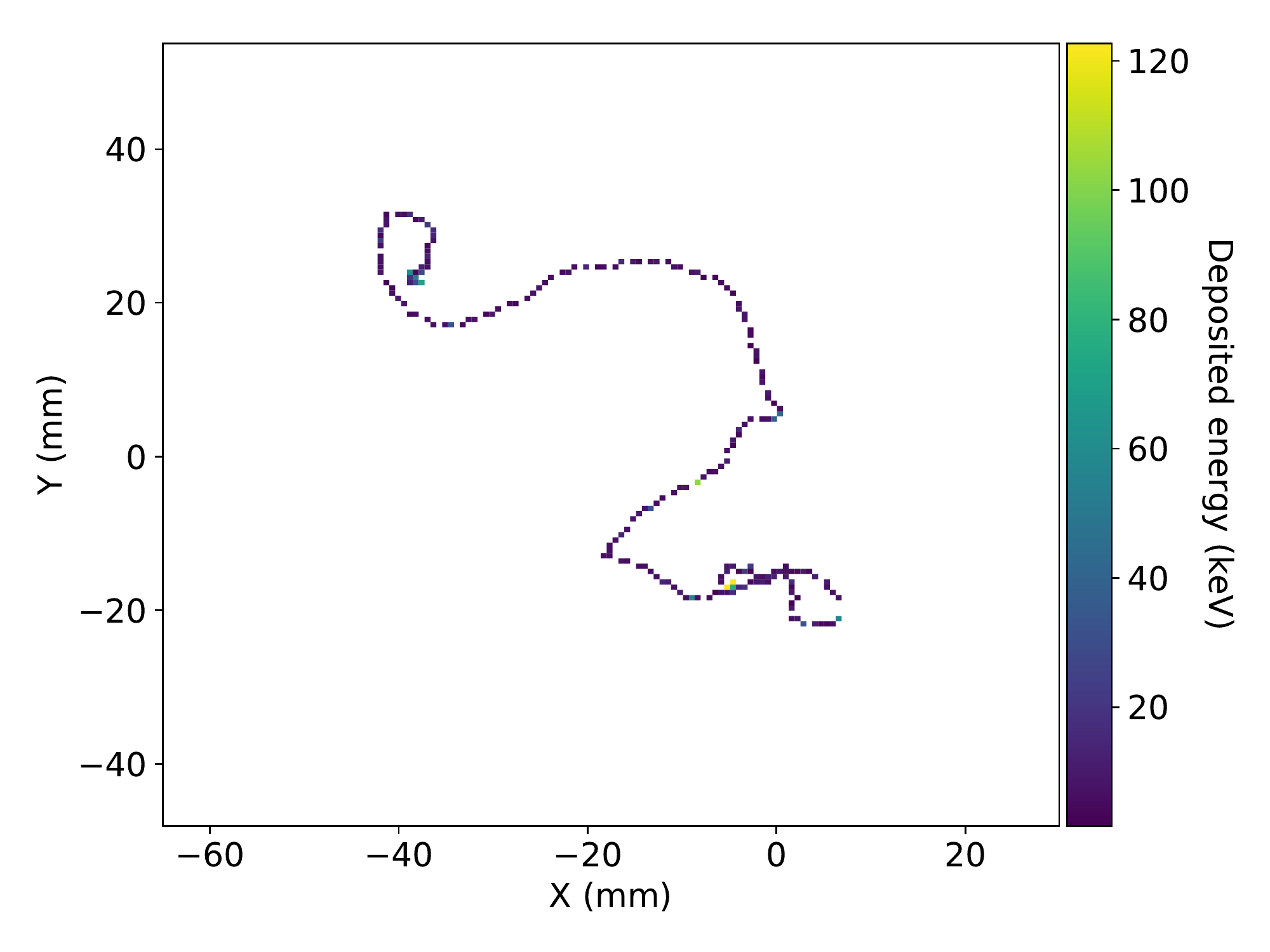}
    \includegraphics[width=0.45\textwidth]{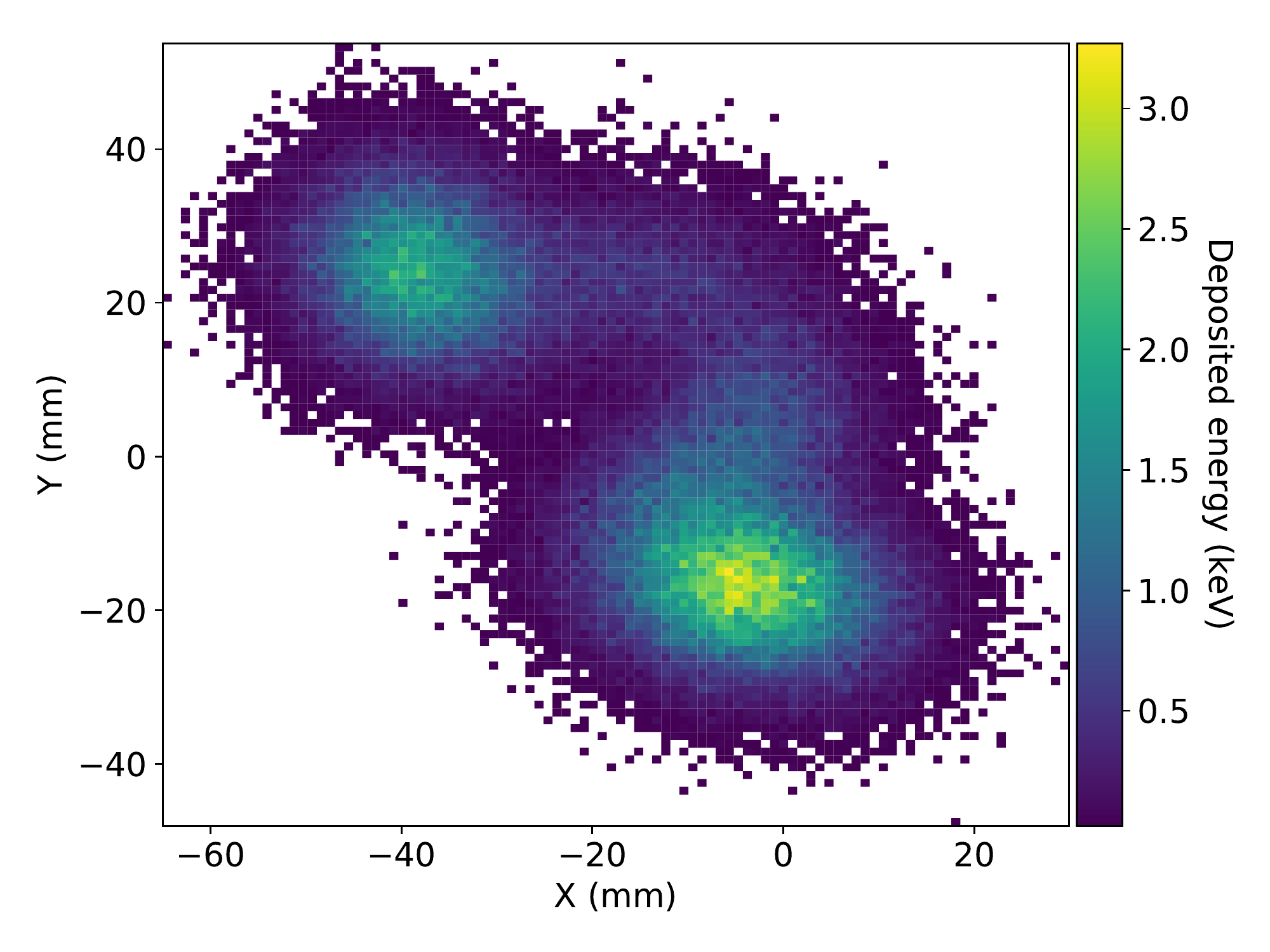}
 \caption{\label{fig:track} Simulated \XE\ \bbonu\ track assuming that no diffusion effects take place during the drift (left), and after drifting 40 cm considering the longitudinal and transverse diffusion coefficients to be 
 \XenonLongitudinalDiffusion\ and \XenonTransverseDiffusion\ respectively (right).}
  \end{center}
\end{figure}


As discussed in \cite{Monrabal:2018xlr}, 
an \HPXeEL\ TPC is an {\em optical TPC}, where both the primary scintillation (\so) and the ionization produce a light signal. In NEXT, the light is detected by two independent sensor planes located behind the anode and cathode. The energy of the event is measured by integrating the amplified EL signal (\st) with a plane of photomultipliers (PMTs). This {\em energy plane} also records the \so\ signal which triggers the start-of-event (\tz).  

Electroluminescent light provides tracking as well, since it is detected a few mm away from production at the anode plane via a denser array of silicon photomultipliers (SiPMs), which constitute the \emph{tracking plane}. As \st\ light is produced in the EL region, near the sensors, \XY\ information transverse to the drift direction can be extracted from the SiPM response. The longitudinal (or \Z) position of the event is inferred from the time difference between the \so\ and \st\ signals (\ie\ the drift time). The drift time will depend on the velocity of the electrons under the influence of an electric field, the so-called drift velocity. Therefore, knowledge of the drift velocity is essential to correctly establish the \Z\ position of the events.

In \NEW\ (and NEXT-100), the \XY\ information comes directly from the light produced by the ionization electrons as they cross the EL gap, after drifting through the TPC active volume. As electrons travel through gas they diffuse, smearing the electron cloud produced by the interacting particle. As a result, the charge distribution arriving at the EL will be spread out. 

In pure xenon, diffusion is large and affects tracking reconstruction. The effect is qualitatively illustrated in \fig\ \ref{fig:track}. A simulated neutrinoless double beta decay (\bbonu) track from \XE\ (\XenonQbb\ deposited energy) interacting in xenon gas at \NewPressure\ is shown on the left panel of the \fig. On the right panel, the same track is imaged after a \SI{40}{cm} drift distance, assuming the known diffusion parameters at that pressure, \ie\ \simto \XenonLongitudinalDiffusion\ and \simto \XenonTransverseDiffusion\ for the longitudinal and transverse diffusion, respectively.
Although the main feature needed to identify the track as two 
electrons (\ie\ the two high-energy depositions at the track end points) can still be discerned, the picture is blurred due to the spreading of the ionization electrons, resulting in loss of information of the particle path. 

The large diffusion characteristic of pure xenon can be mitigated by using a xenon mixture rather than pure xenon. A number of such mixtures showing promising results have been studied by the NEXT collaboration~ \cite{Felkai:2017oeq, Henriques:2017rlj}. In this work we focus instead in a precise measurement of the drift parameters in pure xenon. 

\section {Overview of the \NEW\ detector}
\label{sec.new}

\begin{table}[htp]
\caption{\NEW\ TPC parameters.}
\begin{center}
\begin{tabular}{|c|c|c|c|}
\hline
TPC parameter & Nominal & \RII\ (4734) & \RII\ (4841) \\
\hline
Pressure & \NewPressure & \NewSevenBarPressureRunII & \NewNineBarPressureRunII \\
EL field (E/P) & \NewReducedField & \NewReducedFieldRunII & \NewReducedFieldRunII \\
EL gap & \NewTpcELGap & \NewTpcELGap & \NewTpcELGap \\
$V_{gate}$ & \NewGateVoltageAtFifteenBar & \NewGateVoltageSevenBarRunII & \NewGateVoltageNineBarRunII\\
Length & \NewTpcLength & \NewTpcLength & \NewTpcLength \\
Diameter &  \NewTpcDiameter & \NewTpcDiameter & \NewTpcDiameter \\
Fiducial mass & \NewFiducialMass & \NewFiducialMassSevenBar & \NewFiducialMassNineBar\\
Drift length & \NewTpcDriftLength & \NewTpcDriftLength & \NewTpcDriftLength \\
Drift field & \NewDriftField & \NewDriftField& \NewDriftField \\
$V_{cathode}$ &\NewCathodeVoltageAtFifteenBar &  \NewCathodeVoltageSevenBarRunII & \NewCathodeVoltageNineBarRunII\\
\hline\hline
\end{tabular}
\end{center}
\label{tab.TPC}
\end{table}%

The \NEW\ detector has been thoroughly described elsewhere \cite{Monrabal:2018xlr} and only a brief summary of its main features is offered here. It has three main subsystems, the TPC, the energy plane and the tracking plane. Table \ref{tab.TPC} shows the main parameters of the TPC. The
energy plane is instrumented with \NewNumberOfPMT\ \NewTypePMT\  PMTs located \NewCathodeToPMTs\ behind the cathode, providing a coverage of \NewPMTCoverage. The tracking plane is instrumented with \NewNumberOfSiPM\ SiPMs SensL series-C distributed in a square grid at a pitch of \NewSipmPitch. An ultra-pure \NewBarrelICS-thick copper shell (ICS) acts as a shield in the barrel region. The tracking plane and the energy plane are supported by \NewPlatesICS-thick pure copper plates.
 
The detector operates inside a pressure vessel fabricated with a radiopure titanium alloy (\NewPressureVesselMaterial) surrounded by a lead shield. Since a long electron lifetime is a must, xenon circulates in a gas system where it is continuously purified. The whole setup sits on top of a tramex platform elevated over the ground in HALL-A of LSC. 

\section{Experimental setup}
\label{sec.krdata}

\subsection* {Krypton calibrations in the \NEW\ detector}
\label{sec.krcal}

\begin{figure}[tbh!]
  \begin{center}
    \includegraphics[width=0.6\textwidth]{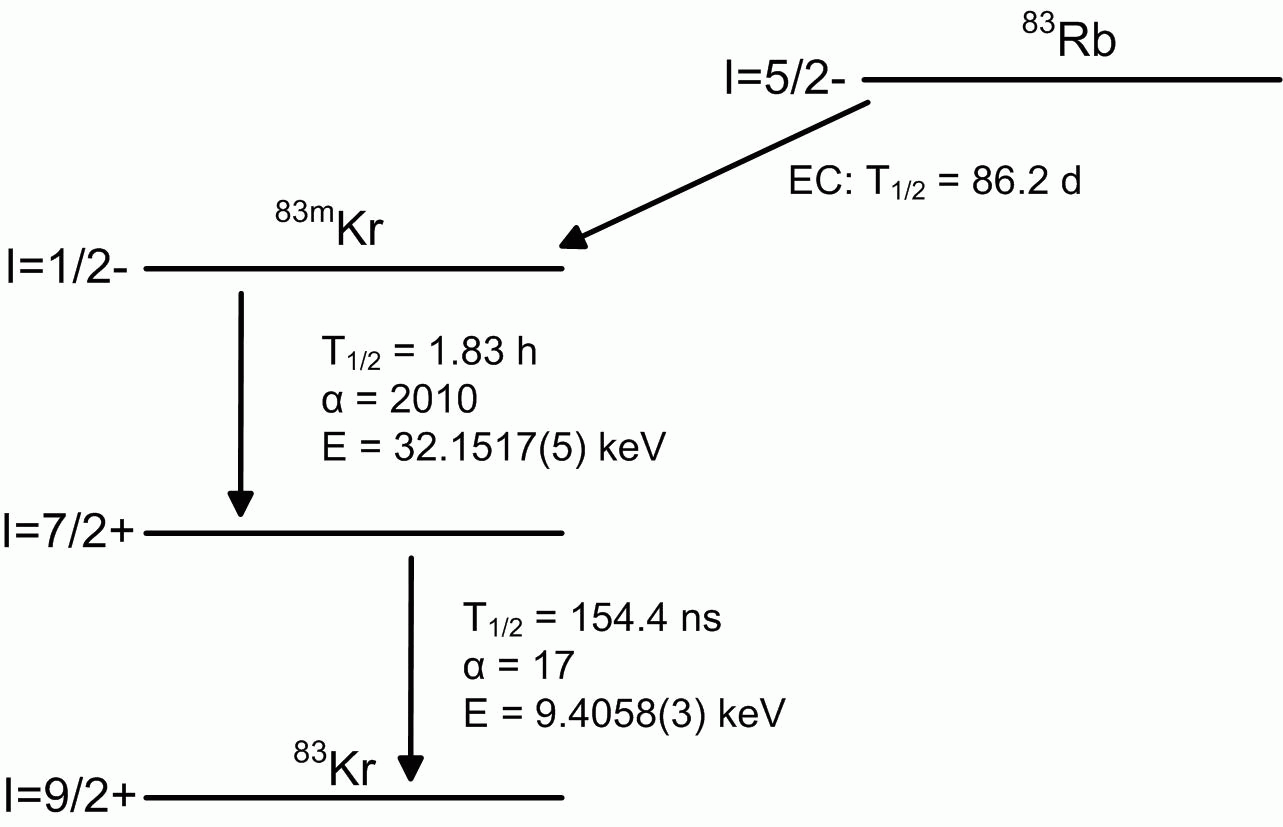} \hfill
    \caption{\label{fig:Rb_decay_scheme} \Rb{83} decay scheme.}
  \end{center}
\end{figure}

Figure \ref{fig:Rb_decay_scheme} shows the decay scheme of a \Rb{83} nucleus. The exotic rubidium isotope decays to \Kr{83m} via electron capture with a lifetime of \RbLifetime. Krypton then decays to the ground state via two consecutive electron conversions releasing \KrEnergyI\ and \KrEnergyII, respectively. The decay rate is dominated by the first conversion with a half-life of \KrLifetime\ (the second one has a very short half-life of \KrLifetimeShort). The total released energy is \KrEnergy\ and the ground state of \Kr{83} is stable.

The rubidium source is a sample of small (a few mm diameter) porous zeolite balls, stored in a dedicated section of the gas system. \Kr{83m} nuclei produced after the electron conversion of \Rb{83} emanate from the zeolite and flow with the gas inside the chamber, spreading uniformly through the detection volume and producing point-like deposits of monochromatic energy (given the short half life of the second electron conversion, the measured energy of \Kr{83m} decays is \KrEnergy). The  source has an intensity of \RbIntensity. The rate of \Kr{83m} decays is limited by the data acquisition to a comfortable value of about  \KrRate. In addition, given the short half-life of \Kr{83m} the krypton activity is reduced to negligible levels a few hours after removing the rubidium source. 
\par

\subsection*{Krypton data periods for \RII}
\label{sec.krp}

Krypton data for \RII\ were taken in two different periods. The first period ranged between \KryptonSevenBarStartRunII\ and \KryptonSevenBarEndRunII, at an operational pressure of \NewSevenBarPressureRunII. The second one occurred between \KryptonNineBarStartRunII\ and \KryptonNineBarEndRunII, at \NewNineBarPressureRunII. 
\par

\subsection*{Datasets}

\begin{table}[!htb]
\caption{\label{tab:krRunConditions}Detector operating conditions for each analyzed data set.}
\begin{center}
\begin{tabular}{cccc}
\Xhline{1.2pt}
Reduced drift field   & Pressure                   & Temperature 			  	& $V_{cathode}$ \\
(V $\cdot$ cm$^{-1}$ $\cdot$ bar$^{-1}$) &   (bar)  & ($^{\circ}$C) & (kV) \\ \Xhline{1.2pt}
53.64 $\pm$ 0.33  & 7.178 $\pm$ 0.018 &	22.286 $\pm$ 0.009	& 28 \\                        	
48.55 $\pm$ 0.30  & 7.175 $\pm$ 0.018 &  22.242 $\pm$ 0.007 	& 26 \\                       	
43.46 $\pm$ 0.27  & 7.172 $\pm$ 0.018 &	22.253 $\pm$ 0.004	& 24 \\                       	
38.30 $\pm$ 0.22  & 7.181 $\pm$ 0.015 &	22.261 $\pm$ 0.003	& 22 \\                      	
33.18 $\pm$ 0.16  & 7.182 $\pm$ 0.009 &	22.267	$\pm$	0.003	& 20 \\
28.08 $\pm$ 0.14  & 7.182 $\pm$ 0.008 &	22.271 $\pm$ 0.002	& 18 \\                     	    
22.97 $\pm$ 0.11  & 7.182 $\pm$ 0.008 &	22.275 $\pm$	0.002	& 16 \\ \Xhline{1.2pt}
42.03 $\pm$ 0.25  & 9.065 $\pm$ 0.019 &	22.160 $\pm$ 0.019  & 29.5 \\
38.03 $\pm$ 0.22  & 9.064 $\pm$ 0.017 &	22.293 $\pm$ 0.004	& 27.5 \\                       	
34.02 $\pm$ 0.20  & 9.066 $\pm$ 0.018 &	22.273 $\pm$ 0.005	& 25.5 \\                       	
30.02 $\pm$ 0.17  & 9.066 $\pm$ 0.018 &	22.263 $\pm$ 0.006	& 23.5 \\   
26.02 $\pm$ 0.15  & 9.065 $\pm$ 0.015 &	22.251 $\pm$ 0.006	& 21.5 \\ \Xhline{1.2pt}
\end{tabular}
\end{center}
\end{table}

As both drift velocity and diffusion depend on the drift field, seven (five) runs with different cathode voltages, ranging from \NewCathodeVoltageKrDiffSevenBarMinRunII\ to \NewCathodeVoltageKrDiffSevenBarMaxRunII\ (\NewCathodeVoltageKrDiffNineBarMinRunII - \NewCathodeVoltageKrDiffNineBarMaxRunII) were taken for the \NewSevenBarPressureRunII\ (\NewNineBarPressureRunII) period. The EL voltage was fixed at \NewGateVoltageSevenBarRunII\ and \NewGateVoltageNineBarRunII\ throughout the two periods, respectively. Detailed information on the detector for each reduced drift field considered is shown in \tab\ \ref{tab:krRunConditions}, with the reduced drift field defined as:

\begin{equation}
\frac{E^*}{P} = \frac{E}{P} \frac{\TPT}{\TOX} \frac{\ZP}{\ZPOX} 
\end{equation}

\noindent with \E\ being the drift field in V$\cdot$cm$^{-1}$, \PR\ the pressure in bar, \TPT\ the detector temperature in K, \TOX = \NewTo\ a reference temperature to normalize the field, \ZP\ the compressibility factor at the operational pressure 
(\NewSevenBarPz\ for \NewSevenBarPressureRunII, \NewNineBarPz\ for \NewNineBarPressureRunII\ \cite{doi:10.1021/ie50467a036}) and \ZPOX\ a reference compressibility factor to normalize the field, considered to be \NewPo\ for the reference temperature \TOX\ and 1 bar pressure.

In the following, we refer to the conditions showed in the first row of \tab\ \ref{tab:krRunConditions} as the standard conditions of the detector, that is, a pressure of \NewSevenBarPressureRunII\ and a reduced drift field of \NewDriftFieldRunII. Indeed, most of the \NEW\ data taken during 2017 were obtained in these conditions.

The associated error on both the pressure and temperature corresponds to half the difference between the maximum and minimum values registered during each run. The pressure values, measured by a pressure gauge connected to the vessel, were stored every 30 seconds. The temperature values were stored every 10 minutes and measured by temperature sensors in the tracking plane boards. On the other hand, the error in the drift field is given by the systematic error in the drift distance, that is, the distance between cathode and the start of the EL region (\NewTpcDriftLength).


%
In addition to these data, simulated \Kr{83m} events have been analyzed to cross-check the procedures developed for the analysis. The data were generated using NEXUS, the simulation framework developed by the NEXT Collaboration and based on Geant4 \cite{Agostinelli:2002hh}. 

\section{Selection of krypton candidates and event reconstruction}
\label{sec:sel}

\begin{figure}[tbh!]
  \begin{center}
    \includegraphics[width=0.6\textwidth]{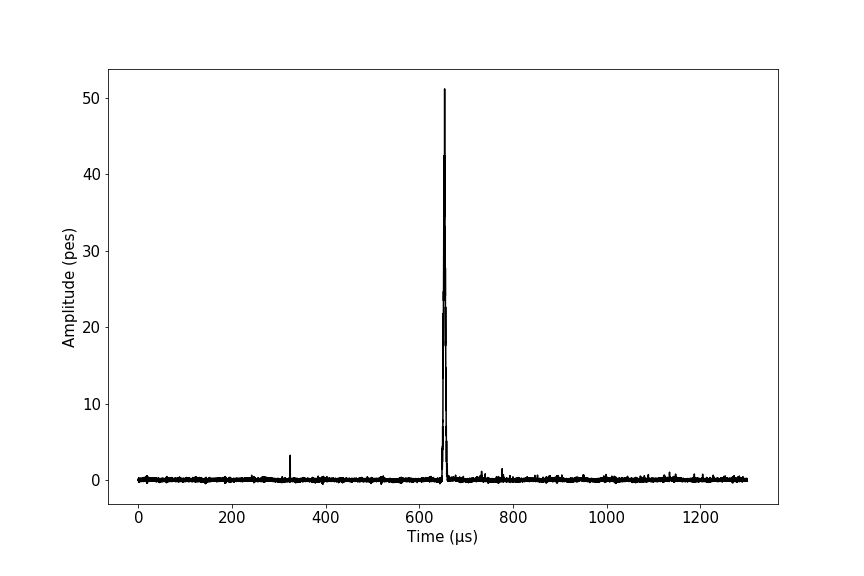} \hfill 
    \caption{\label{fig:cwf} A krypton calibrated waveform. The primary scintillation, \so, can be easily seen at the beginning of the waveform followed by the secondary scintillation, \st, located at the center of the window.}
  \end{center}
\end{figure}

We refer to \cite{Martinez-Lema:2018ibw} for a detailed description of the data selection and event reconstruction. 
The detector triggers on the krypton \st\ signals using the two central PMTs of the energy plane. The recorded data are PMT and SiPM waveforms. 
As described in \cite{Monrabal:2018xlr}, the PMT waveforms show a negative swing due to the effect of the front-end electronics. The first step in the processing is to apply a deconvolution algorithm \cite{Monrabal:2018xlr, Alvarez:2018xuc}, to the negative-swing, arbitrary-baseline, uncalibrated raw-waveforms (RWFs), to produce positive-only, zero-baseline, calibrated waveforms (CWFs). Figure \ref{fig:cwf} shows the CWF waveform corresponding to the sum of the PMTs in the energy plane. The detector triggers in the \st\ signal, centered in the middle of the data acquisition window. The \so\ signal appears in the left of the waveform, indicating that the event has drifted most of the chamber

The first half of the CWF (\ie\ drift time less than \KrSearchWindow) is processed by a peak-finding algorithm tuned to find small signals.  A single \so\ is identified in about half of the events. 
The second half of the CWF (\ie\ drift time more than \KrSearchWindow) is then processed by the same peak finding algorithm, this time tuned to find larger signals. Most of the time a single \st\ candidate is found. Only events with exactly one \so\ and one \st\ are accepted for the analysis. Furthermore, a positive signal in at least one sensor of the tracking plane for the same time window of the \st\ is required. A SiPM is considered to have signal if the total charge in that time window exceeds \KrSiPMEminRunII\ (photoelectrons). This threshold was taken to reduce the impact of SiPM dark noise at the operational temperature of the detector. 

Next, the position of the selected events is reconstructed. 
The \Z\ coordinate of the events is computed multiplying the drift time (obtained as the difference
between \so\ and \st) by the drift velocity. The transverse coordinates \XY\ of the event are obtained using the position and integrated charge of the SiPMs. Concretely, the \XY\ position is computed as the barycenter, the charge-weighted position of the SiPMs.

Finally, additional selections are applied to further increase the purity of the \Kr{83m} dataset. First, a \KrFidVolumeRRunII\ radial cut is applied to select a fiducial region away from the edges of the chamber where solid angle effects and fringe fields are important. Second, an energy selection is made. 

\begin{figure}[!htb]
  \begin{center}
    \includegraphics[align=t, height=0.5\textwidth]{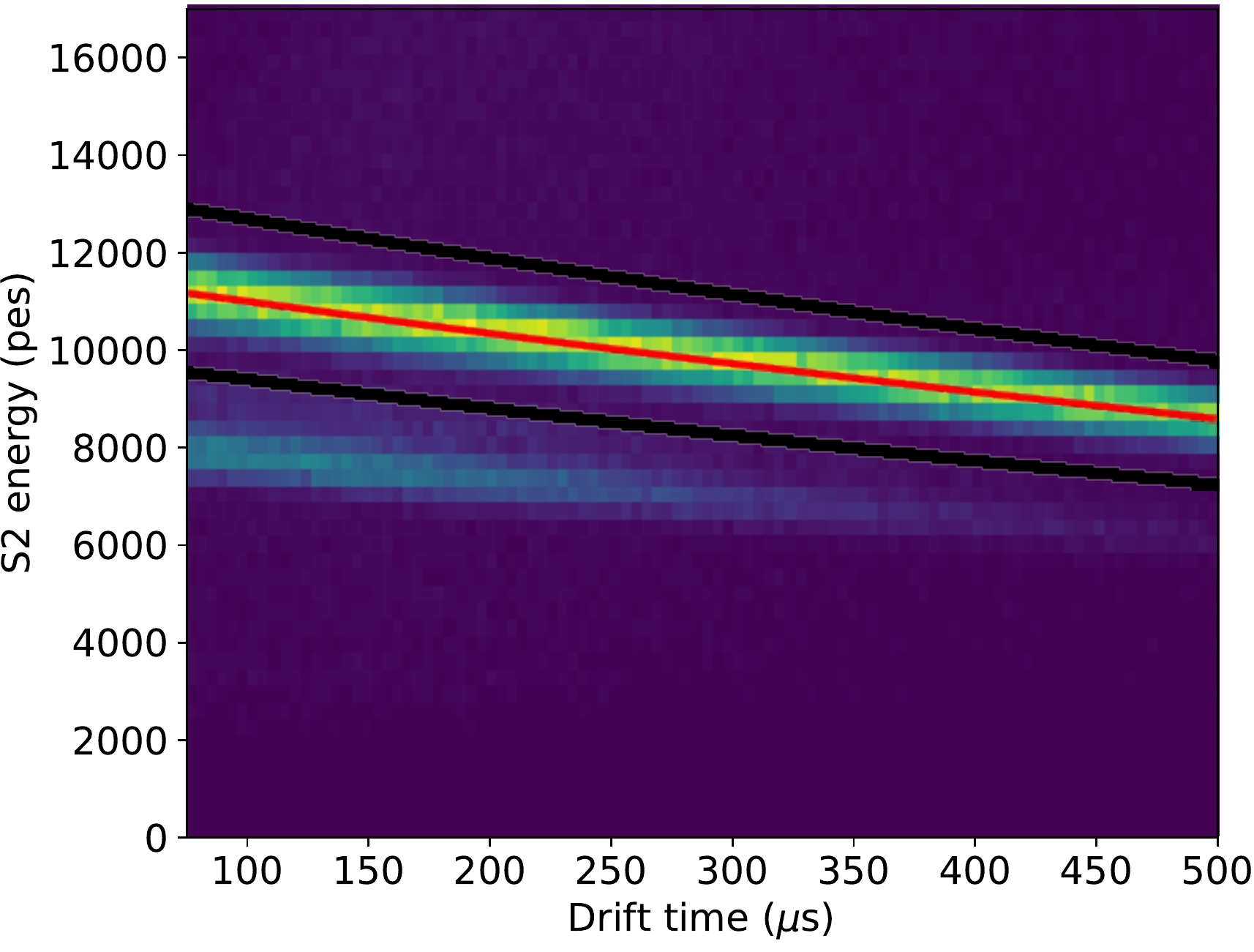}
    \caption{\label{fig:lifetime} Energy of the \st\ charge as a function of the drift time for a reduced drift field of \NewDriftFieldRunII during the \NewSevenBarPressureRunII\ period. An exponential fit (in red) to the distribution provides a measurement of the lifetime of 
  \SI{1617 +- 40}{\micro\second}.}
  \end{center}
\end{figure}

\begin{figure}[!htb]
  \begin{center}
	\includegraphics[align=t, height=0.5\textwidth]{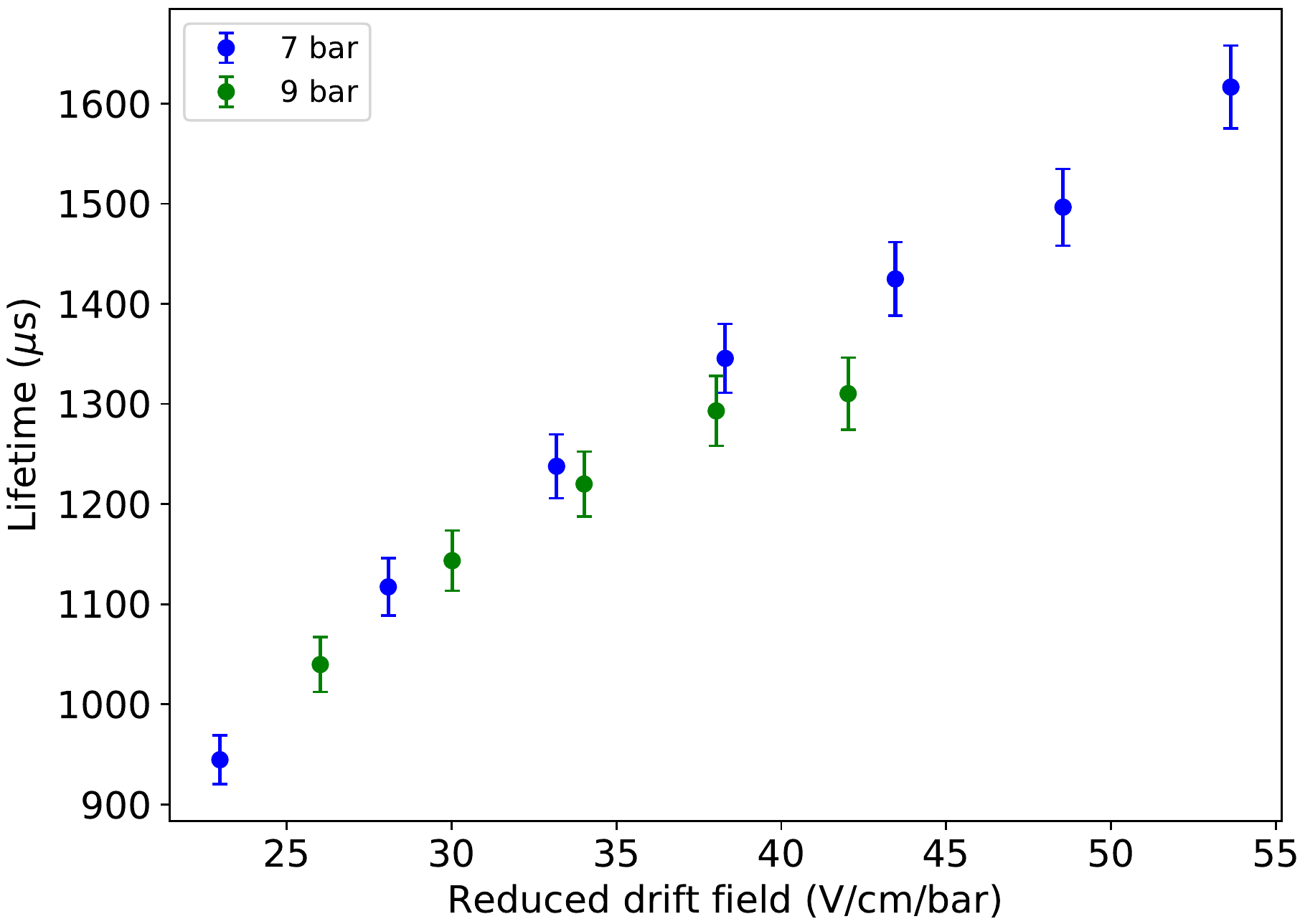}
    \caption{\label{fig:krLifetime} Measured lifetimes for each reduced drift field data set, in blue (green) the \NewSevenBarPressureRunII\ (\NewNineBarPressureRunII) data.}
  \end{center}
\end{figure}

For the energy selection, the lifetime of each run is measured by fitting the distribution of \st\ energy as a function of the drift time,
as illustrated in figure \ref{fig:lifetime}. The lifetime has to be measured for each run as it depends not only on the attachment but also on the drift field \cite{Gonzalez-Diaz:2017gxo}:

\begin{equation}
\LT = (\eta  \VDR)^{-1}
\end{equation}

\noindent with $\eta$~ being an attachment coefficient and \VDR\ the drift velocity. To measure the lifetime, \LT, an exponential is used as fit model:

\begin{equation}
E = E_{0} e^{-t/\LT}
\end{equation}

\noindent where $E_{0}$~ is the energy at zero drift time. The measured lifetime for each reduced drift field data set is shown in figure~\ref{fig:krLifetime}. 



\begin{figure}[!htb]
  \begin{center}
	\includegraphics[align=t, height=0.5\textwidth]{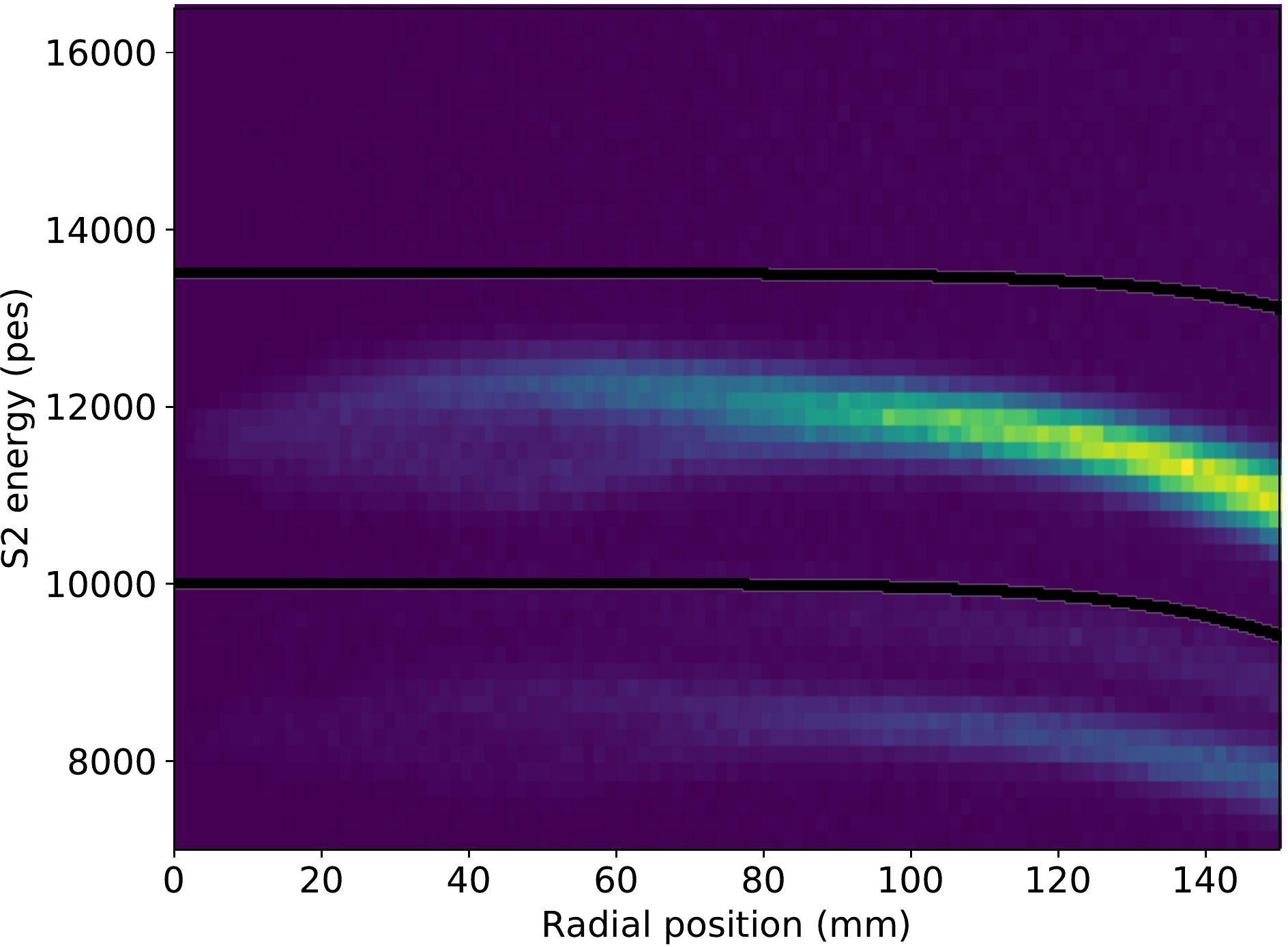}
    \caption{\label{fig:radial} Energy of the \st\ signal corrected by lifetime. Events whose corrected energy is between the black lines pass the energy selection cut.}
  \end{center}
\end{figure}

The \st\ energy corrected by attachment varies across the detector transverse dimensions due to the fact that the solid angle covered by the energy plane depends on the position of the event. Therefore, events have to be selected accordingly, as shown on \fig\ \ref{fig:radial}. After all selections, around 23\% of the triggered events remain. For more details on the technique, the energy calibration of the chamber using \Kr{83m} decays is fully described in [KRYPTON].

\section{Drift velocity}
\label{sec.dv}

\begin{figure}[!htb]
  \begin{center}
    \includegraphics[height=0.4\textwidth]{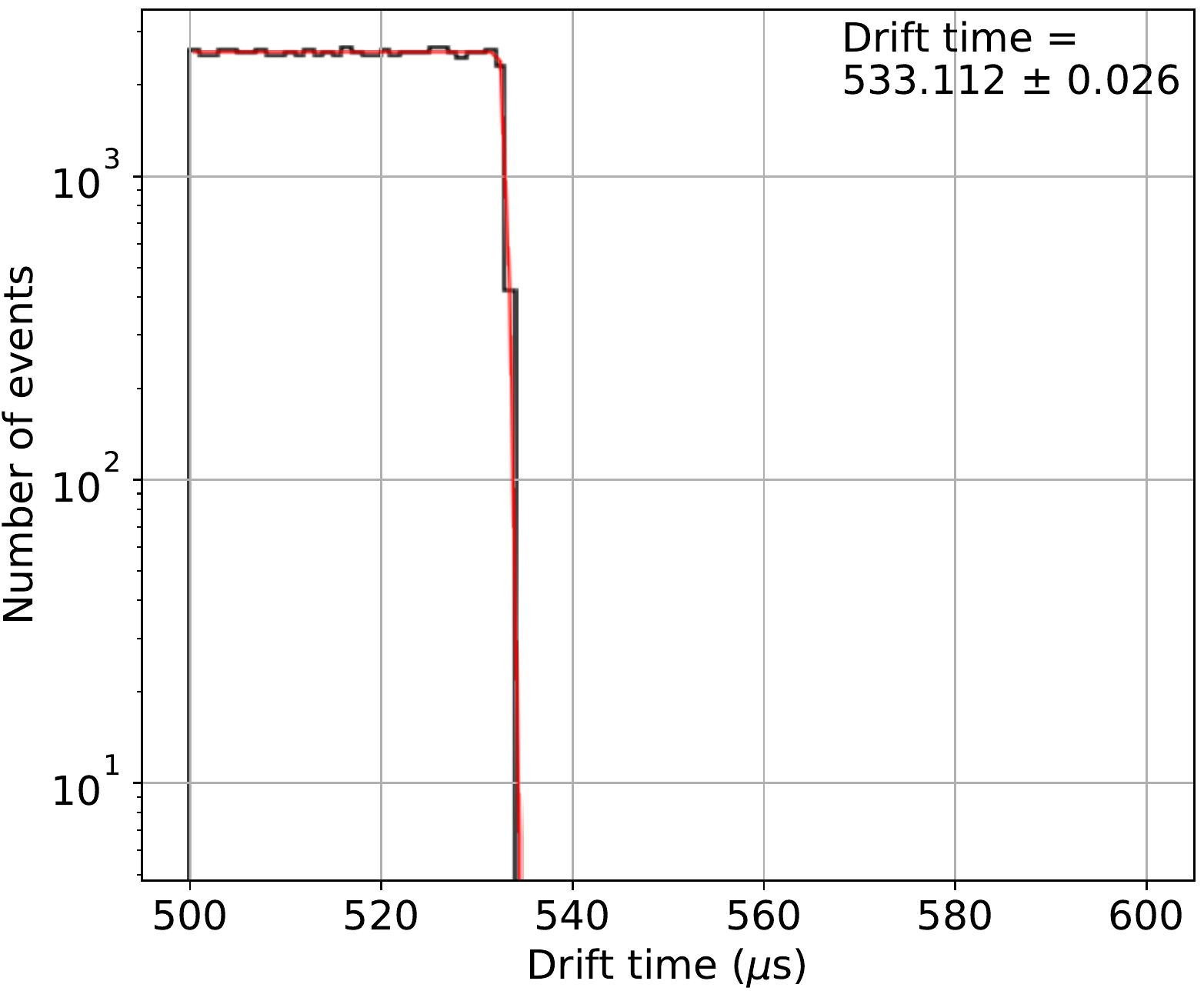}
	\includegraphics[height=0.4\textwidth]{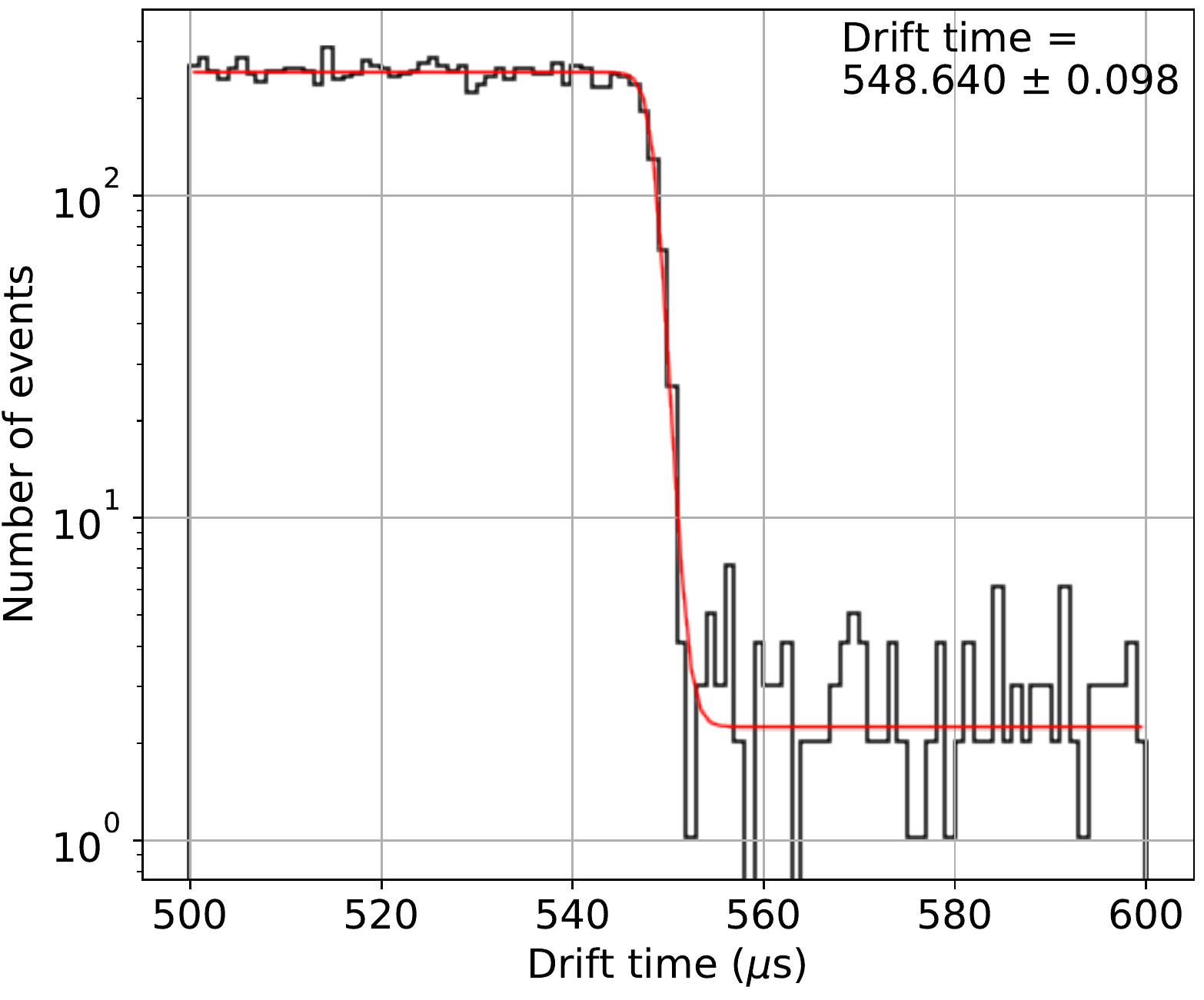}    \caption{\label{fig:driftTime} Drift time distribution end-point for the simulated data (left) and for the Run II standard conditions dataset (right). A Logistic function fit to the distribution yields a maximum drift time of \MaxDriftTimeMC\ and \MaxDriftTimeStandardRunII, for simulated and standard conditions data respectively.}
  \end{center}
\end{figure}

 \begin{figure}[!htb]
  \begin{center}
    \includegraphics[height=0.5\textwidth]{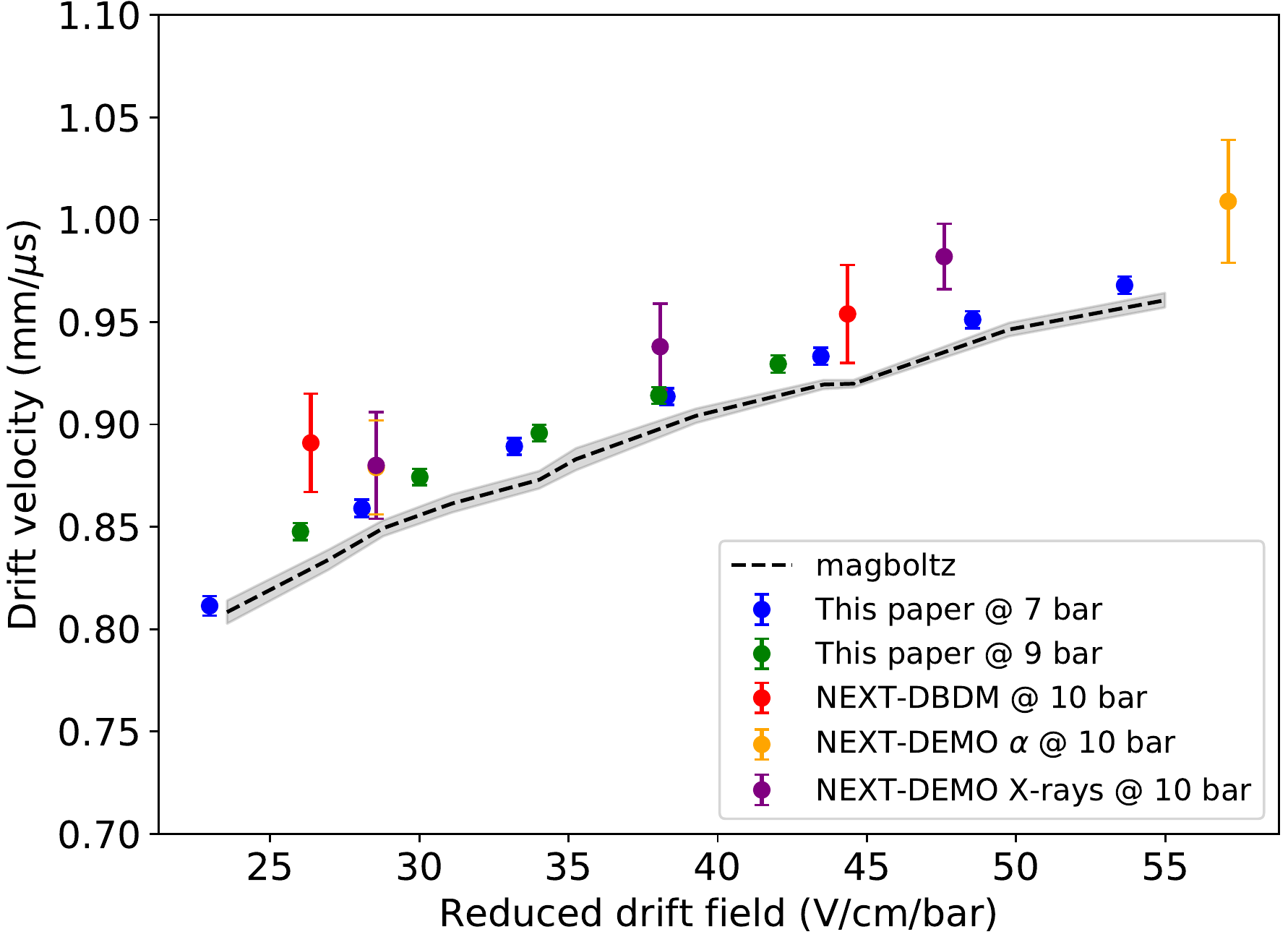}
 \caption{\label{fig:driftVelComparison} Drift velocity dependence on the reduced drift field. The comparison between our measurements (blue and green for 7.2 and 9.1 bar data, respectively) and Magboltz simulations (black) is shown. The figure also shows previous drift velocity results by the NEXT Collaboration. Although the simulated values are systematically below the measurements, all results are in agreement.}
  \end{center}
\end{figure}

The drift velocity can be obtained by measuring the drift time for a known distance. In NEXT, the reference inside the chamber is the cathode plane, as its position is well measured. Since the cathode is the farthest $z$ from the EL region, the maximum drift time will correspond to that position, $Z_{cath}$. Consequently, the drift velocity is immediately given by: 

\begin{equation}
\VDR = \frac{Z_{cath} }{t_{max}}
\label{eq:DriftVel}
\end{equation}

However, a small correction has to be added to \eq\ \ref{eq:DriftVel} to also consider the track time of the electrons crossing the EL region. We define the time positions of the \st\ and \so\ signals as those of the waveform's time bin with maximum energy. Given the point-like nature of the events considered, the electron cloud arriving at the EL will follow a gaussian distribution due to diffusion. Since \st\ scintillation takes place uniformly during the electron cloud transit time throughout the EL region, the resulting \st\ signal can be seen approximately as a convolution with a square function, whose time duration equals the electron transit time ($t_{EL}$). Consequently, the signal peaking time is expected to be shifted by $t_{EL/2}$.

A Magboltz simulation of the drift field at the operational EL voltage (\NewGateVoltageSevenBarRunII\  at \NewSevenBarPressureRunII\ and \NewGateVoltageNineBarRunII\ at \NewNineBarPressureRunII) has been done to estimate the travel time through the EL gap. Doing this, an EL crossing-velocity of \ELVelocitySevenBarRunII\ (\ELVelocityNineBarRunII) was obtained for the \NewSevenBarPressureRunII\ ( \NewNineBarPressureRunII) data. The EL gap in \NEW\ is \NewTpcELGap\, therefore the time it takes to cross half of that distance is $t_{EL/2}$ = \TimeToCrossELSevenBarRunII\ (\TimeToCrossELNineBarRunII). 

On the other hand, the distance between cathode and gate is well known and measured to be $Z_{cath}$ = \NewTpcDriftLength\ and \eq\ \ref{eq:DriftVel} results in:

\begin{equation}
v_d = \frac{\NewTpcDriftLength}{t_{max}  - t_{EL/2}} 
\label{eq:DriftVel_effective}
\end{equation}

The maximum drift time is obtained from the end-point of the drift time distribution, as shown in \fig\ \ref{fig:driftTime}. The number of events, $N$, near the maximum drift time is fitted to a logistic function:

\begin{equation}
N = \frac{A}{1+e^{-k \cdot (t-t_{max})}} + B
\end{equation}

where $A$~ is a constant that accounts for the height of the distribution, $B$~ is related with the number of residual events after the steep drop of the function; $k$~ is related with the steepness of the curve and $t_{max}$~ corresponds to the drift time coordinate at half-height. The parameter $t_{max}$~ gives the distribution end-point and the maximum drift time. 

As a cross-check, Monte Carlo data has been generated with a drift velocity of
\MCDriftVelocity. The fit to the simulated data, showed in figure~\ref{fig:driftTime}, yields a drift velocity for the simulated data of \MCDriftVelocityFitRunII. The central value differs by only 0.21 $\cdot$ 10$^{-3}$~ from the value used for the simulation. This is well below the systematic error considered, of order 0.4\%, which comes from the uncertainty in $Z_{cath}$, $t_{EL/2}$ and $t_{max}$. 

In simulated data the uncertainty of the two first parameters is zero but that is not the case for the latter one, $t_{max}$. The uncertainty on $t_{max}$ has been considered to be equal to the time resolution of the detector, \SI{1}{\micro\second}. Since the systematic error is at least an order of magnitude higher than the difference between measurement and simulation, we conclude that the analysis method does not add significant uncertainties to the result.

The drift velocities for the considered drift field conditions are shown in figure~\ref{fig:driftVelComparison}, where they are also compared with Magboltz simulation results.  For the Run II standard conditions the result of the fit corresponds to a drift velocity of \DriftVelocityStandardRunII. A similar value was obtained in an independent analysis using alpha events in taken with \NEW\ during the Run II \cite{Novella:2018ewv}.The simulations have been run at 1 bar for each considered reduced drift field. The results of the \NewSevenBarPressureRunII\ and \NewNineBarPressureRunII\ data match very well and show a smooth trend. This validates the usefulness of the reduced drift field as a quantity to be used to quote drift velocity results. In this way, results can be easily extrapolated to other operating pressures. 

Our drift velocity results are systematically slightly higher than Magboltz predictions by a mean value of 1.2\%, with the maximum deviation being 1.8\% and the minimum 0.4\%. This small discrepancy between Magboltz simulations and our results is also present in previous measurements of the NEXT Collaboration \cite{Alvarez:2012hu, Lorca:2014sra}, also shown in the figure, and could be due to the fact that the drift field is no longer uniform near the cathode mesh as it highly increases.

\section{Diffusion}
\label{sec.diff}
Diffusion of electrons drifting in gas due to the effect of an electric field follows a gaussian distribution whose standard deviation (RMS), $\sigma_{L,T}$, is defined by:

\begin{equation}
\sigma_{L,T}^2 = \sigma_{0_{L,T}}^2 + 2\cdot D_{L,T} \cdot t_{drift} = \sigma_{0_{L,T}}^2 + 2\cdot D_{L,T} \cdot (t - t_{EL/2})
\label{eq:Diffusion}
\end{equation}

where $D_{L,T}$~ is the longitudinal (L) or transverse (T) diffusion coefficient to be extracted, $\sigma_{0_{L,T}}$~ is a free parameter of the fit that accounts for any additional spread, and $t_{drift}$~ is the drift time of the events. As before, $t_{drift}$~ is the measured drift time $t_{S2}-t_{S1}$~ minus half of the transit time inside the EL region, $t_{EL/2}$. The spread $\sigma_{L,T}$~ in \eq\ \ref{eq:Diffusion} is given in length units, so it follows that $D_{L,T}$ is usually expressed in cm$^2$ $\cdot$ s$^{-1}$ units.

Following \eq\ \ref{eq:Diffusion}, a linear fit to the dependence of the squared spread with drift time is sufficient to obtain $D_{L,T}$. 

\subsection{Longitudinal diffusion}\label{sec:longDiff}

\begin{figure}[!htb]
  \begin{center}
    \includegraphics[height=0.3\textwidth]{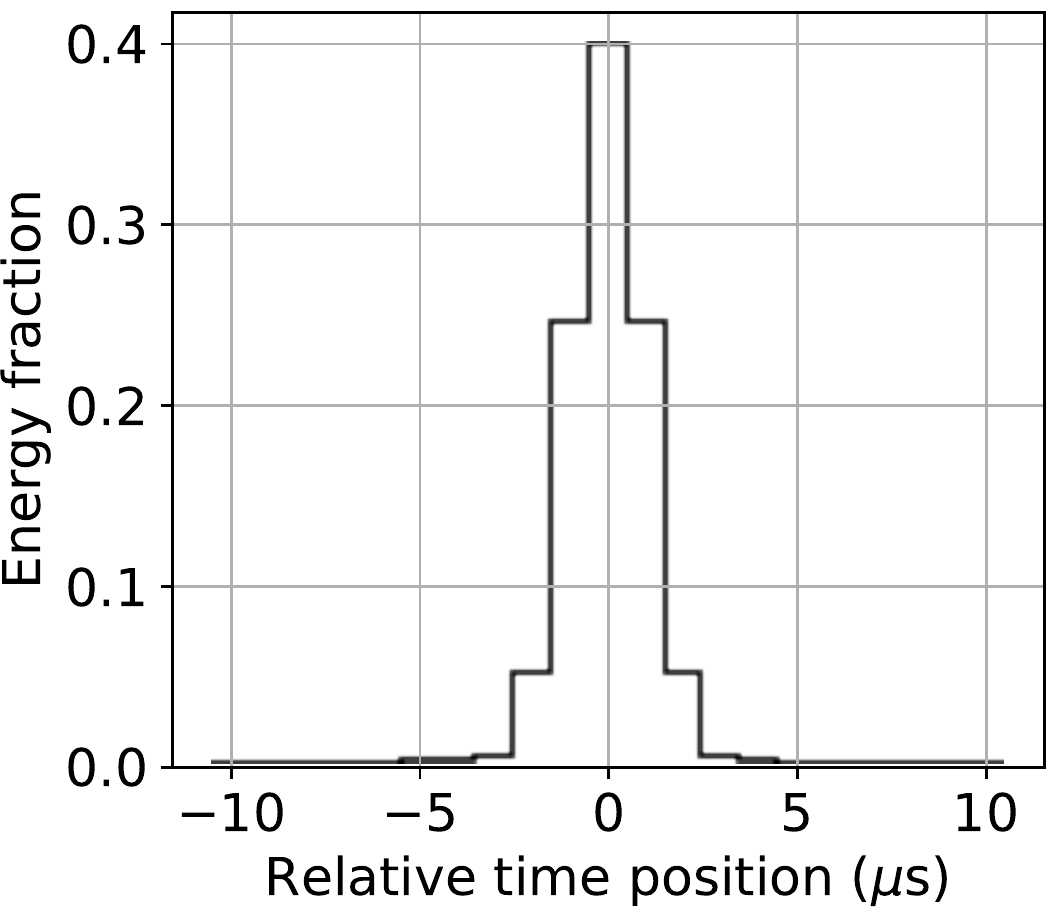}
    \includegraphics[height=0.3\textwidth]{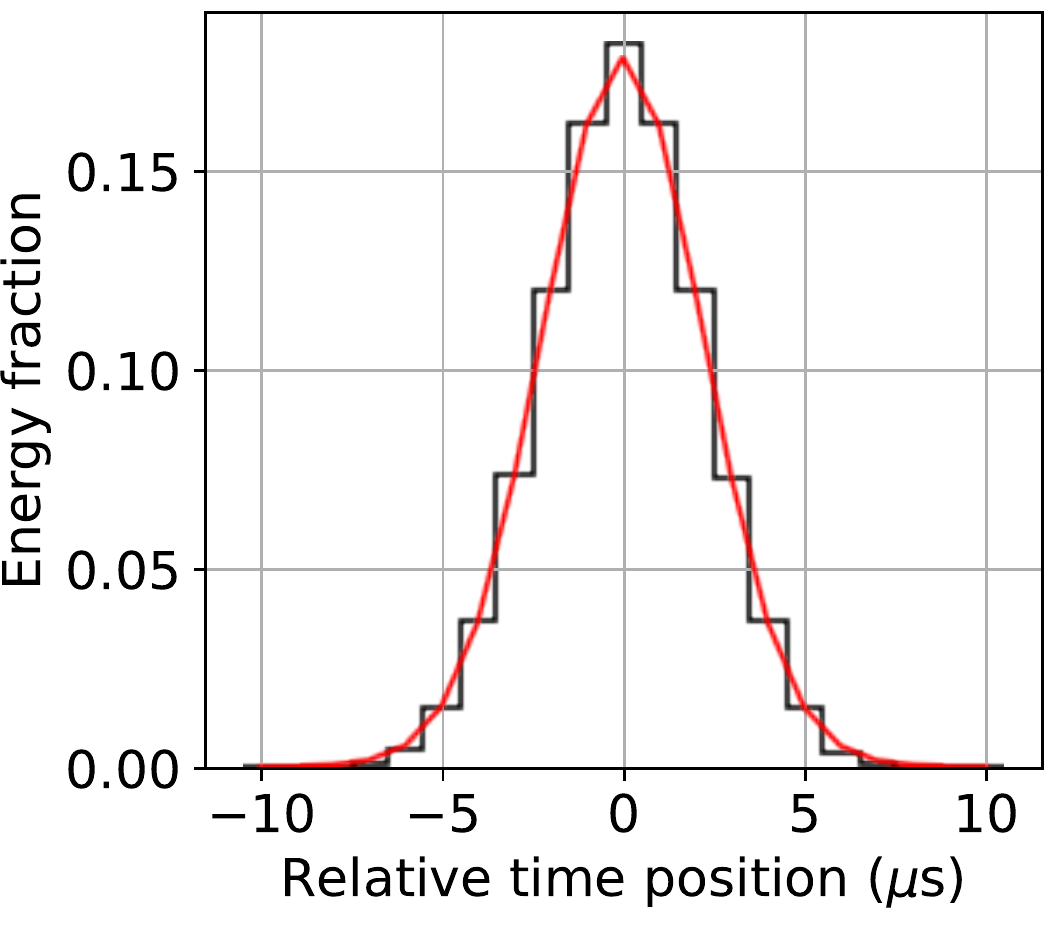}
	\includegraphics[height=0.3\textwidth]{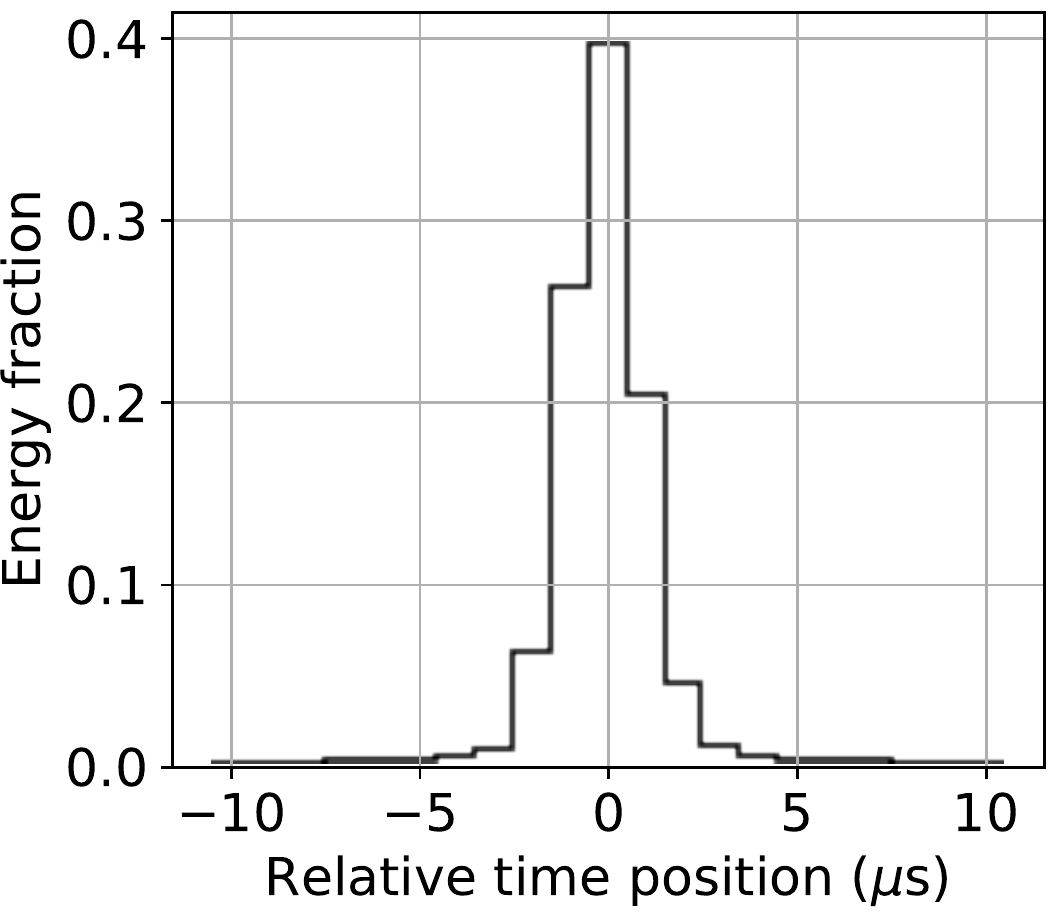}
	\includegraphics[height=0.3\textwidth]{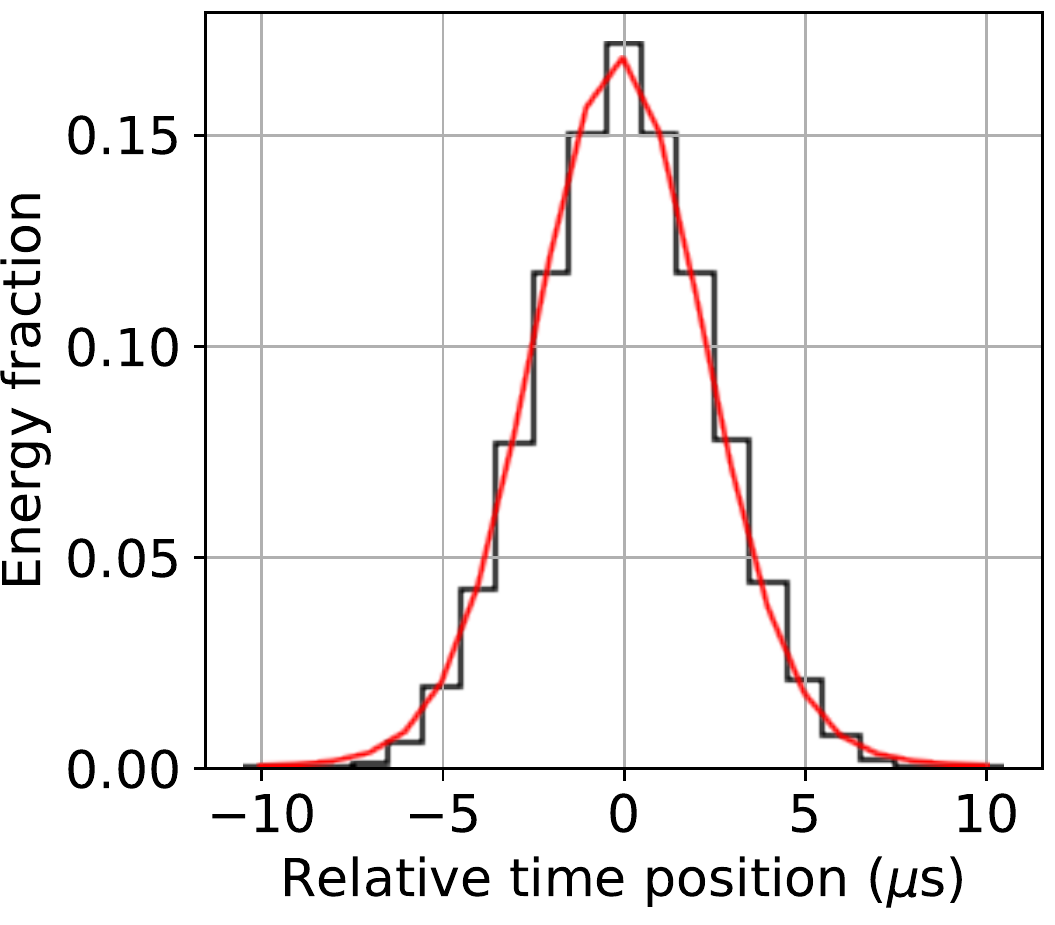}   
	 \caption{\label{fig:waveDiff} Normalized mean waveform for a \SIrange{10}{75}{\micro\second} drift time interval in a region near the EL (left) for simulated data (top) and the standard conditions data (bottom). On the right, the mean waveform for a high (\SIrange{475}{500}{\micro\second}) drift interval is shown. The red curve shows the resulting fit to a convolution of the near-EL waveform with a gaussian of $\sigma$ equal to \SI{2.0}{\micro\second} (simulated data) and \SI{2.1}{\micro\second} (standard conditions data).}
  \end{center}
\end{figure}

The measurement of the longitudinal position is directly related with the drift time. Thus, equation~\ref{eq:Diffusion} now is:

\begin{equation}
\sigma_{L} = \sigma_t \cdot v_d \longrightarrow \sigma_t^2 = \sigma_{0_t}^2 + 2\frac{D_{L}}{v_d^2} t_{drift}
\label{eq:LongDiffusion}
\end{equation}

To measure the longitudinal spread, the detector is divided in drift time intervals and the mean \st\ waveform of each interval computed. To this end, waveforms are normalized to the total energy, and then shifted so that their maximum is centered at $t$=0. Finally the waveforms are averaged to obtain the mean \st\ waveform.

The shape of the \st\ waveforms is due to the convolution of two effects: the diffusion and the effect of the electrons crossing the EL region and the associated light emission. At very low drift distance, near the EL, the \st\ shape only corresponds to the effect of the EL (because electrons have almost not diffused before reaching the EL mesh) while \st\ shapes for events at longer distances are spread out due to both the EL-effect and the diffusion. In other words, the \st\ shapes are the result of a convolution of the distribution at low drift with a gaussian induced by diffusion. A fit to this convolution can be used to obtain the diffusion contribution within each time interval.

The mean waveform measured for the first time bin, at low drift time, is used as the EL-spread distribution for the convolution. Hereafter the drift time range is chosen to be between 10 and \SI{75}{\micro\second}. The lower limit is chosen to avoid events with false \so\ positives, caused by a waveform oscillation at the beginning of the \st\ signal. The higher limit is chosen to have a sufficient range from the statistical point of view, minimizing statistical fluctuations. The rest of the data is divided in \SI{25}{\micro\second} intervals starting at drift times corresponding to \SI{200}{\micro\second} and up to \SI{500}{\micro\second}. Each of the slices is fitted with the convolution described above. The starting point is set at \SI{200}{\micro\second} so that the distributions are dominated by diffusion. An example of a reference waveform and a fit for high drift times are shown in figure \ref{fig:waveDiff} for both simulated and real data.

\begin{figure}[!htb]
  \begin{center}
    \includegraphics[height=0.3\textwidth]{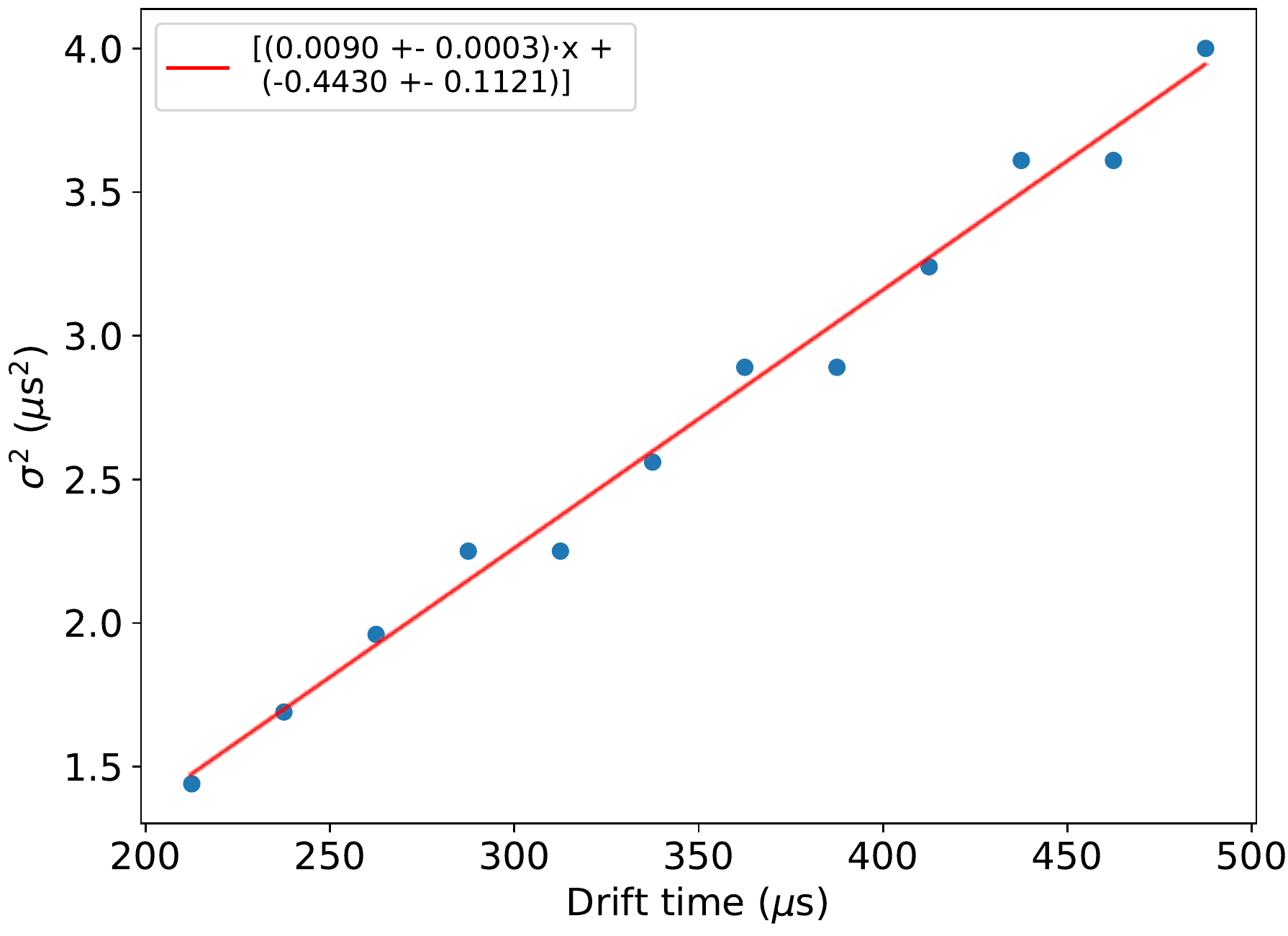}
    \includegraphics[height=0.3\textwidth]{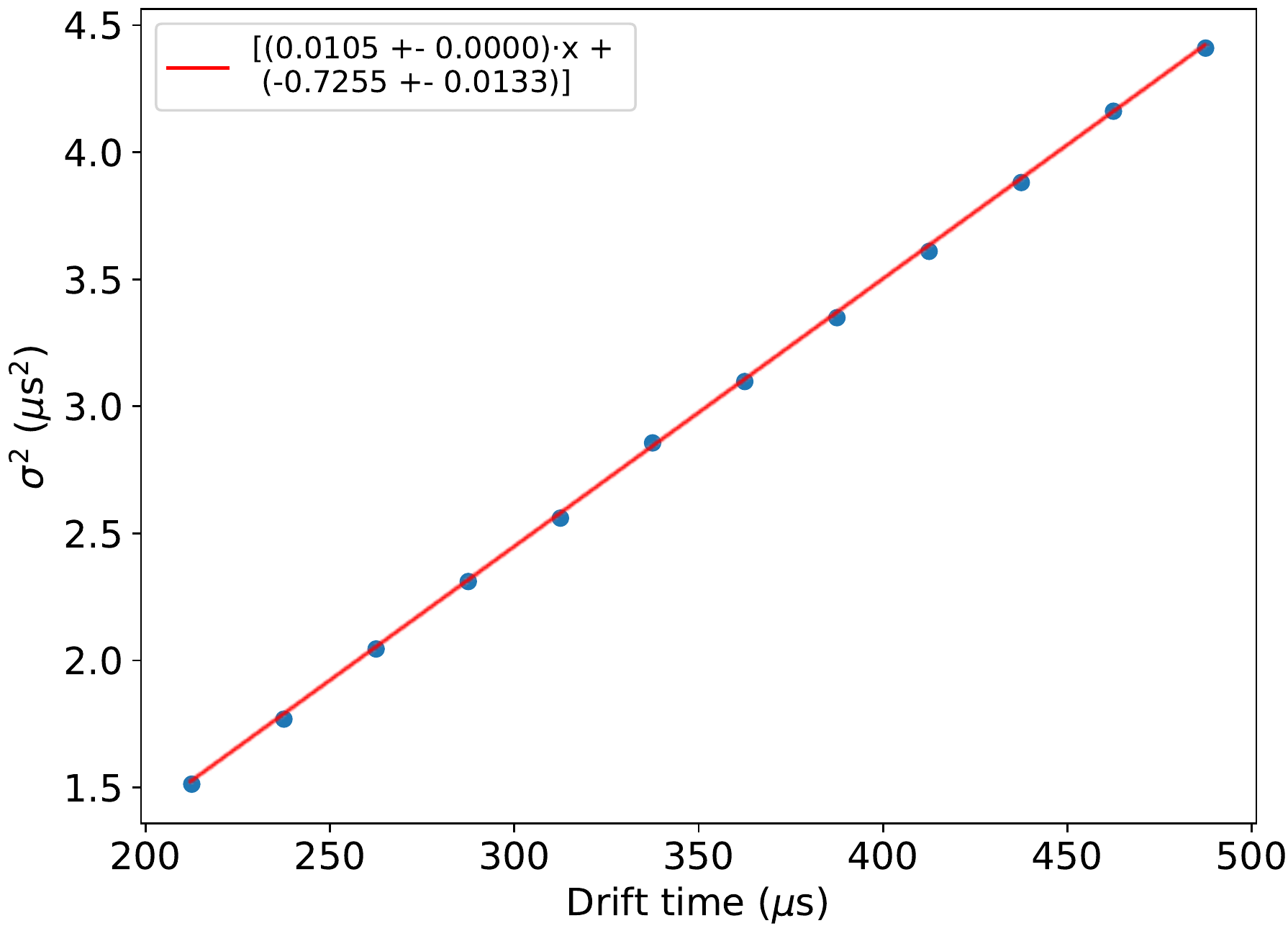}
    \caption{\label{fig:waveFit}Longitudinal spread squared, estimated using the mean waveform method, as a function of drift time. The red line shows the linear fit to the data. Left: simulation data. Right: standard conditions data.}
  \end{center}
\end{figure}

The square of the obtained gaussian spread for each drift time interval versus the center of the drift time interval is fitted to a linear function to obtain $D_L$, according to equation~\ref{eq:LongDiffusion}, as shown in figure~\ref{fig:waveFit}. The longitudinal diffusion coefficients for each reduced drift field are shown in table ~\ref{tab:krLongDiff}. The coefficients have been extrapolated to normal temperature, \AmbientNormalTemperature, and transformed into a most common representation with the following expression \cite{Alvarez:2013kqa}:

\begin{equation}
D_L^* = \sqrt{\frac{T^0}{T} \frac{\ZPOX}{\ZP} \frac{2\cdot P\cdot D_L}{v_d}}
\label{eq:generalDiffusion}
\end{equation}
 
where \VDR\ is the drift velocity, $T^0$~ is the temperature we are extrapolating to (\StandardNormalTemperature), \ZPOX~ is the corresponding compressibility factor, $D_L$~ is the diffusion coefficient and $T$, $P$ and \ZP~ are the temperature, pressure and compressibility factor at which $D_L$~ has been measured, respectively. The obtained diffusion can easily be extrapolated to any pressure using \eq\ \ref{eq:generalDiffusion}. This transformation is also valid for the transverse diffusion, discussed in section~\ref{sec:transDiff}.

\begin{table}[!htb]
\caption{\label{tab:krLongDiff}Longitudinal diffusion obtained with the two different approaches discussed in the text.}
\begin{center}
\begin{tabular}{cccc}
\Xhline{1.2pt}
& Reduced drift field                  & \multicolumn{2}{c}{Longitudinal Diffusion ($\sqrt{\textrm{bar}}$ $\cdot$ \micro m/$\sqrt{\textrm{cm}}$)} \\
& (V $\cdot$ cm$^{-1}$ $\cdot$ bar$^{-1}$) & Mean waveform                               & Event-by-event                              \\ \Xhline{1.2pt}

\multirow{7}{*}{\rotatebox[origin=c]{90}{7 bar data}} 
& 53.64 $\pm$ 0.33        & 866.5 $\pm$ 0.5 (stat.) $\pm$ 54.7 (syst.)              & 853.6 $\pm$ 0.3 (stat.) $\pm$ 40.9 (syst.)          \\ 
& 48.55 $\pm$ 0.30        & 900.9 $\pm$ 0.8 (stat.) $\pm$ 56.7 (syst.)              & 882.0 $\pm$ 0.4 (stat.) $\pm$ 42.2 (syst.)          \\
& 43.46 $\pm$ 0.27        & 937.9 $\pm$ 0.8 (stat.) $\pm$ 58.9 (syst.)              & 916.9 $\pm$ 0.5 (stat.) $\pm$ 43.8 (syst.)          \\
& 38.30 $\pm$ 0.22        & 985.6 $\pm$ 0.9 (stat.) $\pm$ 61.4 (syst.)              & 967.3 $\pm$ 0.6 (stat.) $\pm$ 45.6 (syst.)         \\
& 33.18 $\pm$ 0.16        & 1055.1 $\pm$ 1.1 (stat.) $\pm$ 64.8 (syst.)            & 1025.4 $\pm$ 0.9 (stat.) $\pm$ 47.5 (syst.)         \\
& 28.08 $\pm$ 0.14        & 1160.2 $\pm$ 1.9 (stat.) $\pm$ 70.9 (syst.)            & 1116.5 $\pm$ 1.8 (stat.) $\pm$ 51.5 (syst.)         \\
& 22.97 $\pm$ 0.11        & 1359 $\pm$ 5 (stat.) $\pm$ 82 (syst.)                     & 1297 $\pm$ 4 (stat.) $\pm$ 59 (syst.)               \\  \hline
\multirow{5}{*}{\rotatebox[origin=c]{90}{9 bar data}} 
& 42.03 $\pm$ 0.25  	     & 937.6 $\pm$ 1.0 (stat.) $\pm$ 59.7 (syst.)        	  & 937.7 $\pm$ 0.8 (stat.) $\pm$ 45.5 (syst.)          \\ 
& 38.03 $\pm$ 0.22        & 978.4 $\pm$ 1.0 (stat.) $\pm$ 62.0 (syst.)       	  	  & 976.2 $\pm$ 0.8 (stat.) $\pm$ 47.0 (syst.)          \\ 
& 34.02 $\pm$ 0.20        & 1028.6 $\pm$ 0.9 (stat.) $\pm$ 65.1 (syst.)            & 1018.4 $\pm$ 0.7 (stat.) $\pm$ 49.0 (syst.)          \\ 
& 30.02 $\pm$ 0.17        & 1103.6 $\pm$ 1.1 (stat.) $\pm$ 69.5 (syst.)            & 1083.5 $\pm$ 1.0 (stat.) $\pm$ 51.9 (syst.)          \\ 
& 26.02 $\pm$ 0.15        & 1211 $\pm$ 2 (stat.) $\pm$ 76 (syst.)                     & 1189.1 $\pm$ 1.8 (stat.) $\pm$ 56.2 (syst.)               \\ \Xhline{1.2pt}
\end{tabular}
\end{center}
\end{table}

The upper limit of the near-anode region, nominally set at \SI{75}{\micro\second}, has been varied to check that its value does not bias our results. To do that, we apply our diffusion extraction method to simulated data and vary this upper limit from \SI{50}{\micro\second} to \SI{200}{\micro\second}, with the lower limit set at its \SI{10}{\micro\second} nominal value. Diffusion results extracted over this range are compatible with each other. We conservatively assign a systematic error on $D_L$~ based on the maximum difference between the diffusion measurements, corresponding to 5.2\%.

\begin{figure}[!htb]
  \begin{center}
    \includegraphics[height=0.3\textwidth]{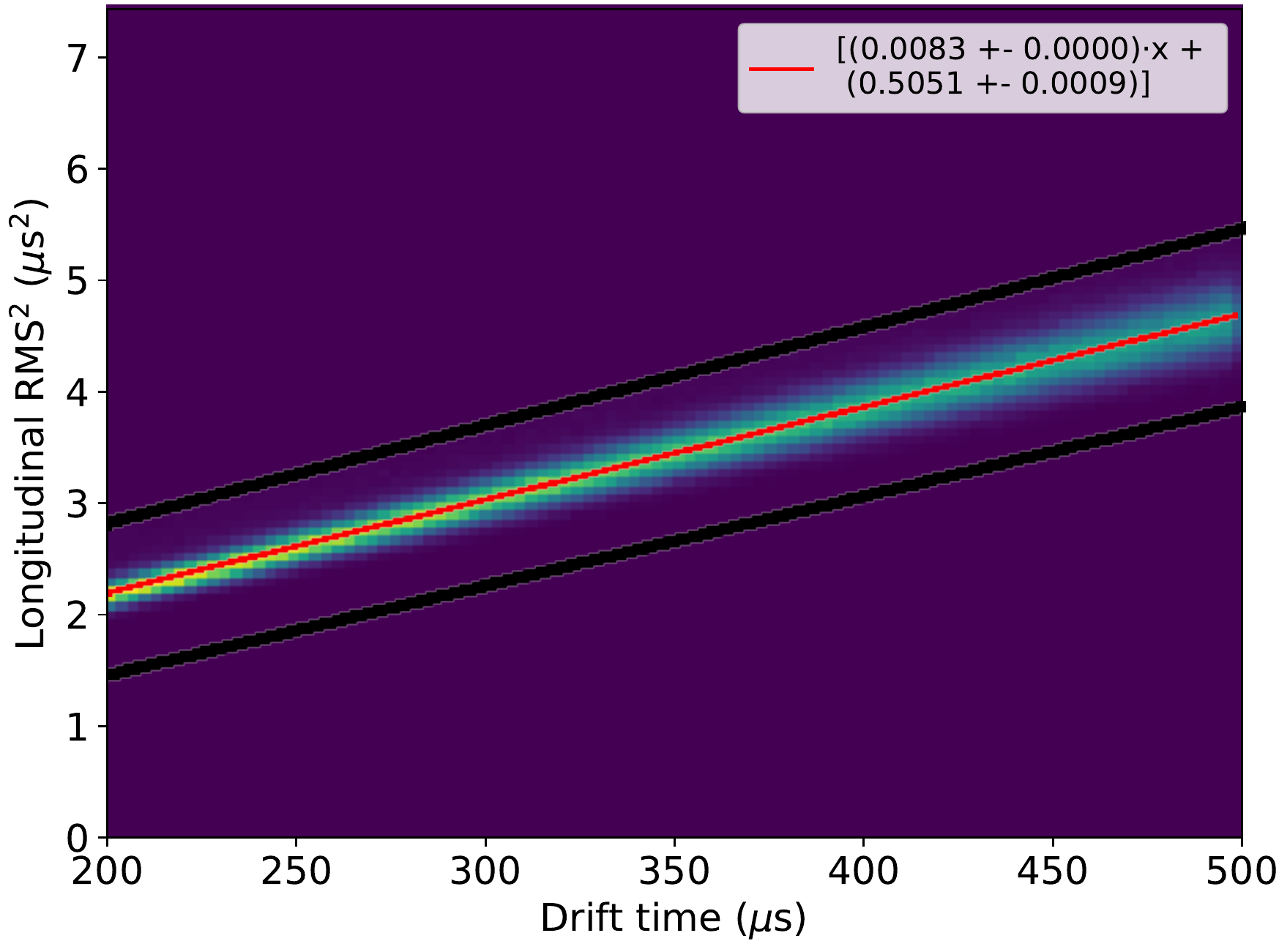}
    \includegraphics[height=0.3\textwidth]{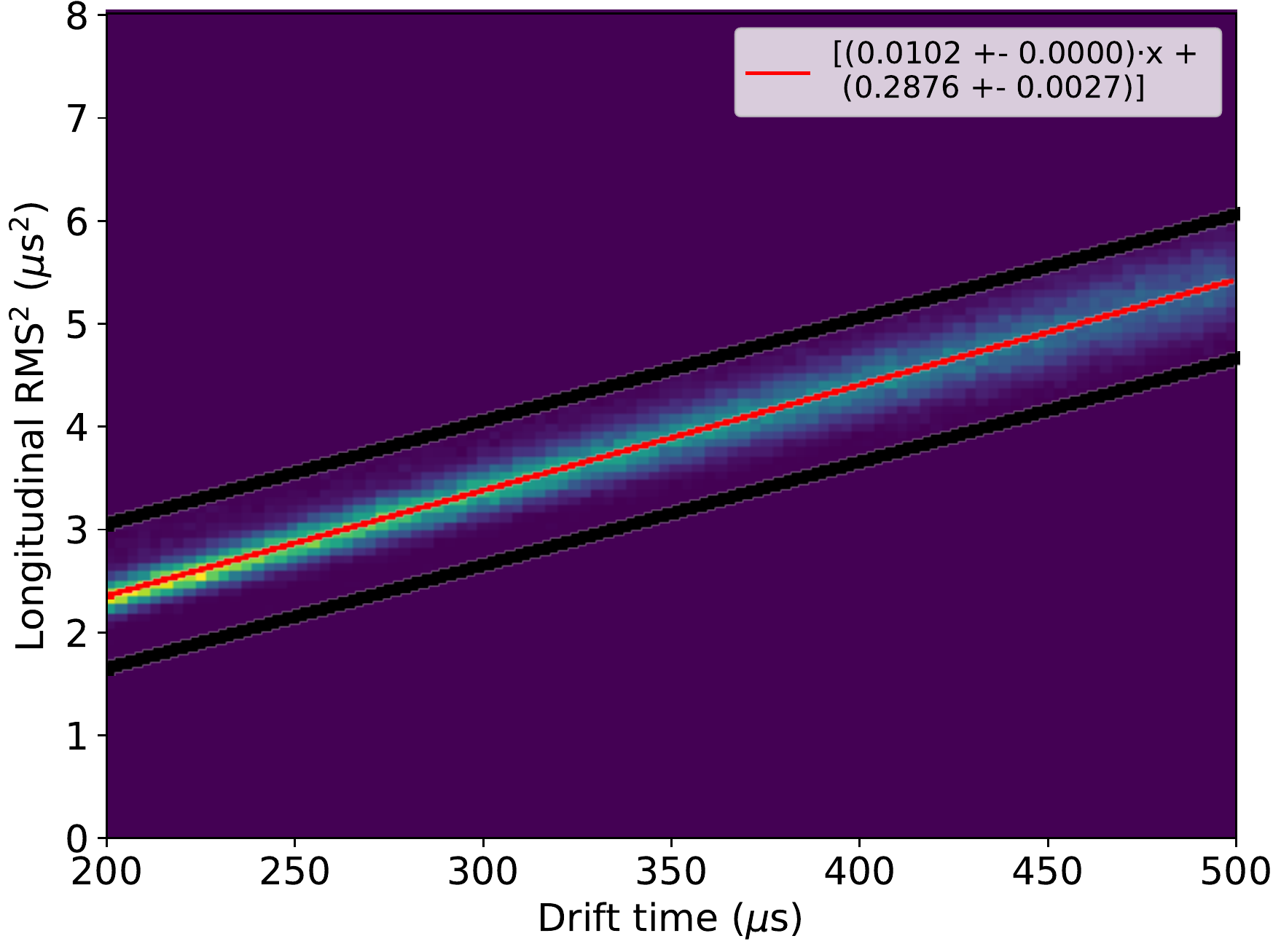}
    \caption{\label{fig:eventDiff}Longitudinal spread squared, estimated on an event-by-event basis via the RMS, as a function of drift time. The red line shows the linear fit to the data. The events considered in the fit are those within the black lines. Left: simulation. Right: Run II standard conditions.}
  \end{center}
\end{figure}

An alternative method to extract $D_L$, based on an event-by-event analysis was studied as well. In this case, the time spread is computed as the RMS of each individual S2 waveform. Shown in figure~\ref{fig:eventDiff}, the distribution of this spread as a function of the drift time of each event is fitted to \eq\ \ref{eq:LongDiffusion}. After applying the same pressure-temperature transformation given in \eq\ \ref{eq:generalDiffusion}, the results shown on table~\ref{tab:krLongDiff} are obtained. As can be seen from the table, the results obtained with the two methods are compatible with each other.

 \begin{figure}[!htb]
  \begin{center}
    \includegraphics[height=0.5\textwidth]{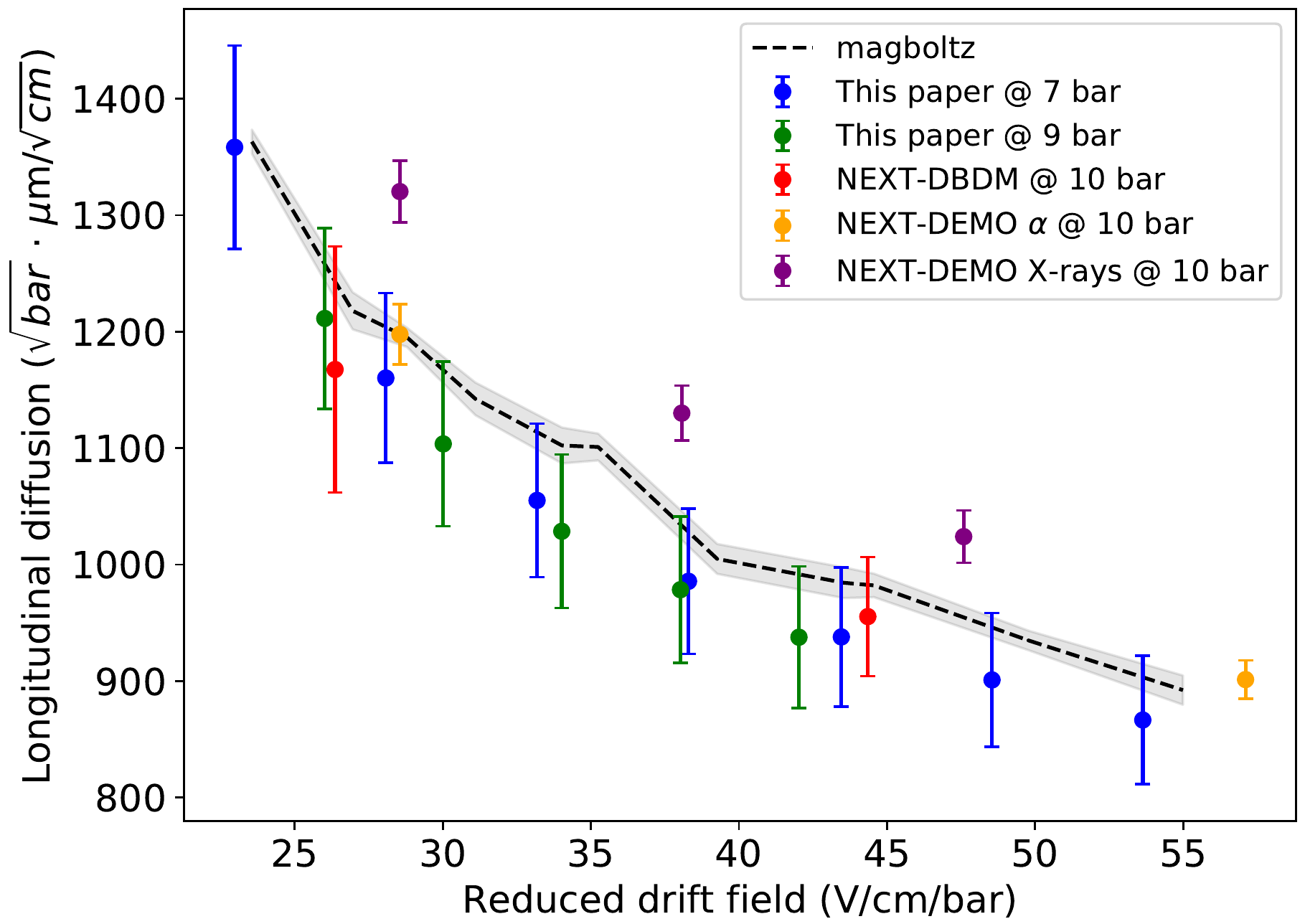}
 \caption{\label{fig:longDiffComparison} Dependence of the reduced longitudinal diffusion ($D_L^*$) with the reduced drift field. Our measurements for both 7.2 (blue) and 9.1 (green) bar data are compatible with Magboltz predictions (black, uncertainties in grey) and previous NEXT collaboration measurements.}
  \end{center}
\end{figure}

For simulated data at the Run II standard conditions (and if ignoring at this point the Z, T, P corrections) the value for $D_L$ in the case of the mean waveform method is [299.6 $\pm$ 2.0 (stat.) $\pm$ 5.5 (syst.)] \micro m/$\sqrt{\textrm{cm}}$. The extracted number matches with great precision the value used for generating the data, 300 \micro m/$\sqrt{\textrm{cm}}$, demonstrating the validity of the approach. For the event-by-event measurement, the result is [288.68 $\pm$ 0.03 (stat.) $\pm$ 0.74 (syst.)] which, again, is close to the diffusion used in the simulation process. The relative difference between the measured and simulated diffusion coefficients has been added as an additional systematic error to the data measurements: 0.13\% for the mean waveform approach, 3.77\% for the event-by-event one.

A comparison of our results with Magboltz simulations run at \AmbientNormalTemperature\ and 1 bar (therefore, density scaling is assumed) for the considered reduced drift fields in this analysis, is shown on figure~\ref{fig:longDiffComparison}. Our results closely match the Magboltz simulations, with a mean deviation of 3.5\%, a maximum deviation of 6.6\% and a minimum deviation of 0.4\%. The agreement validates the pressure scaling (P-scaling) of the longitudinal diffusion given by \eq\ \ref{eq:generalDiffusion} for the range of pressures considered. Furthermore, the new results contained in this work are compatible with previous measurements made by the NEXT collaboration in smaller prototypes at a similar pressure (10 bar) \cite{Alvarez:2012hu, Lorca:2014sra}. These earlier results are also shown in figure~\ref{fig:longDiffComparison}.

\subsection{Transverse diffusion}\label{sec:transDiff}

\begin{figure}[!htb]
  \begin{center}
    \includegraphics[align = c, width=0.4\textwidth]{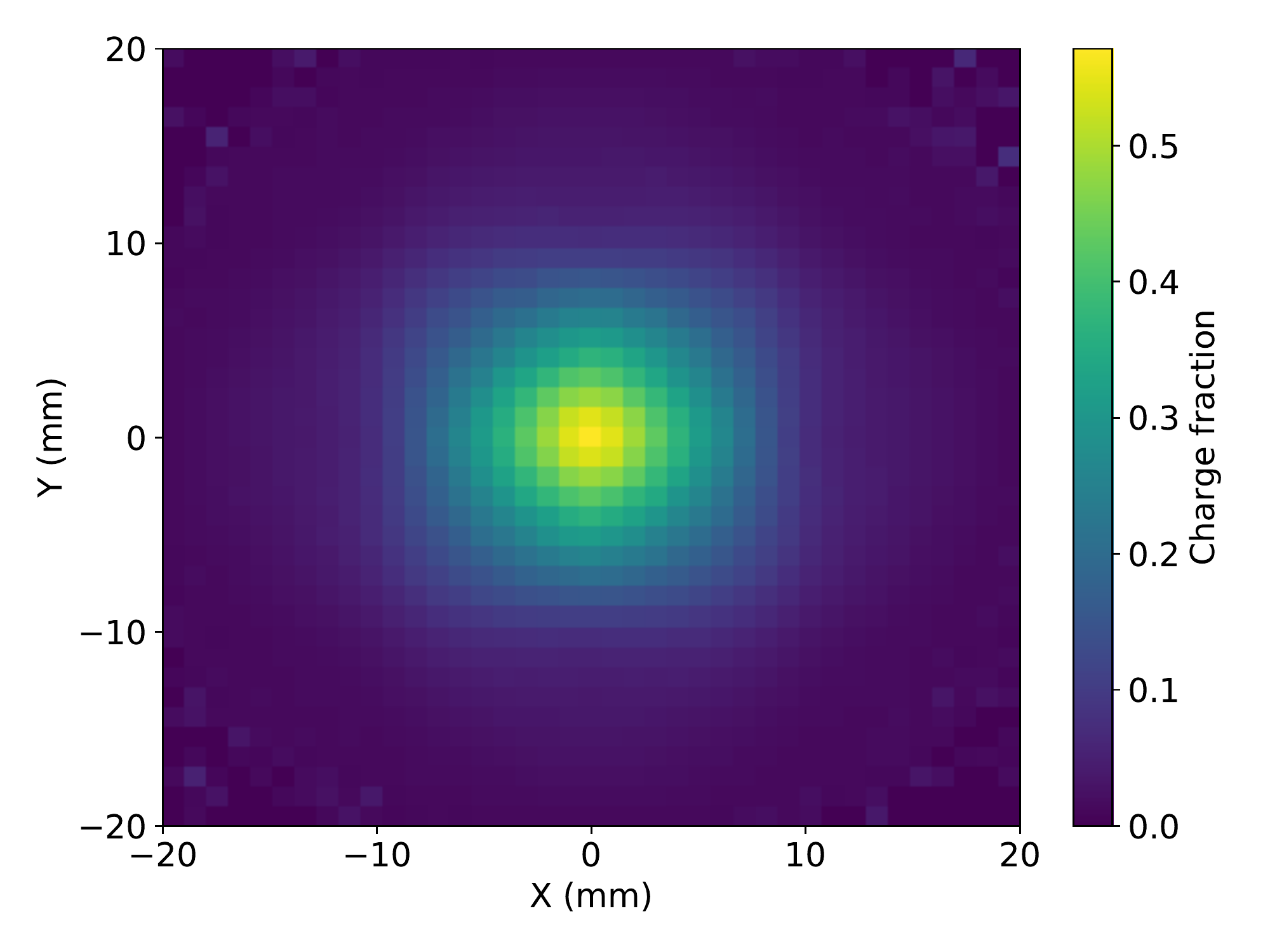}
    \includegraphics[align = c, width=0.4\textwidth]{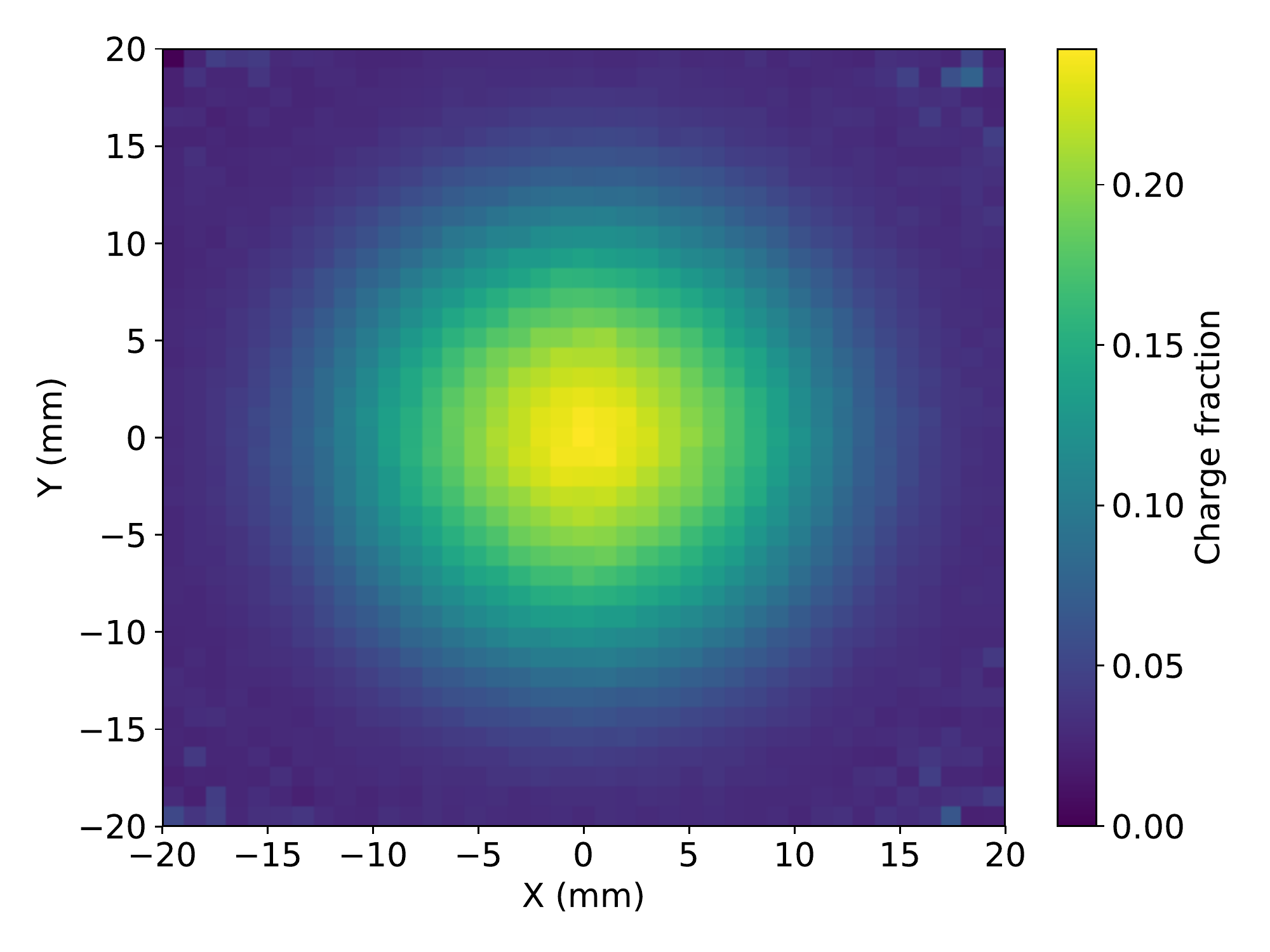}
	\includegraphics[align = c, width=0.4\textwidth]{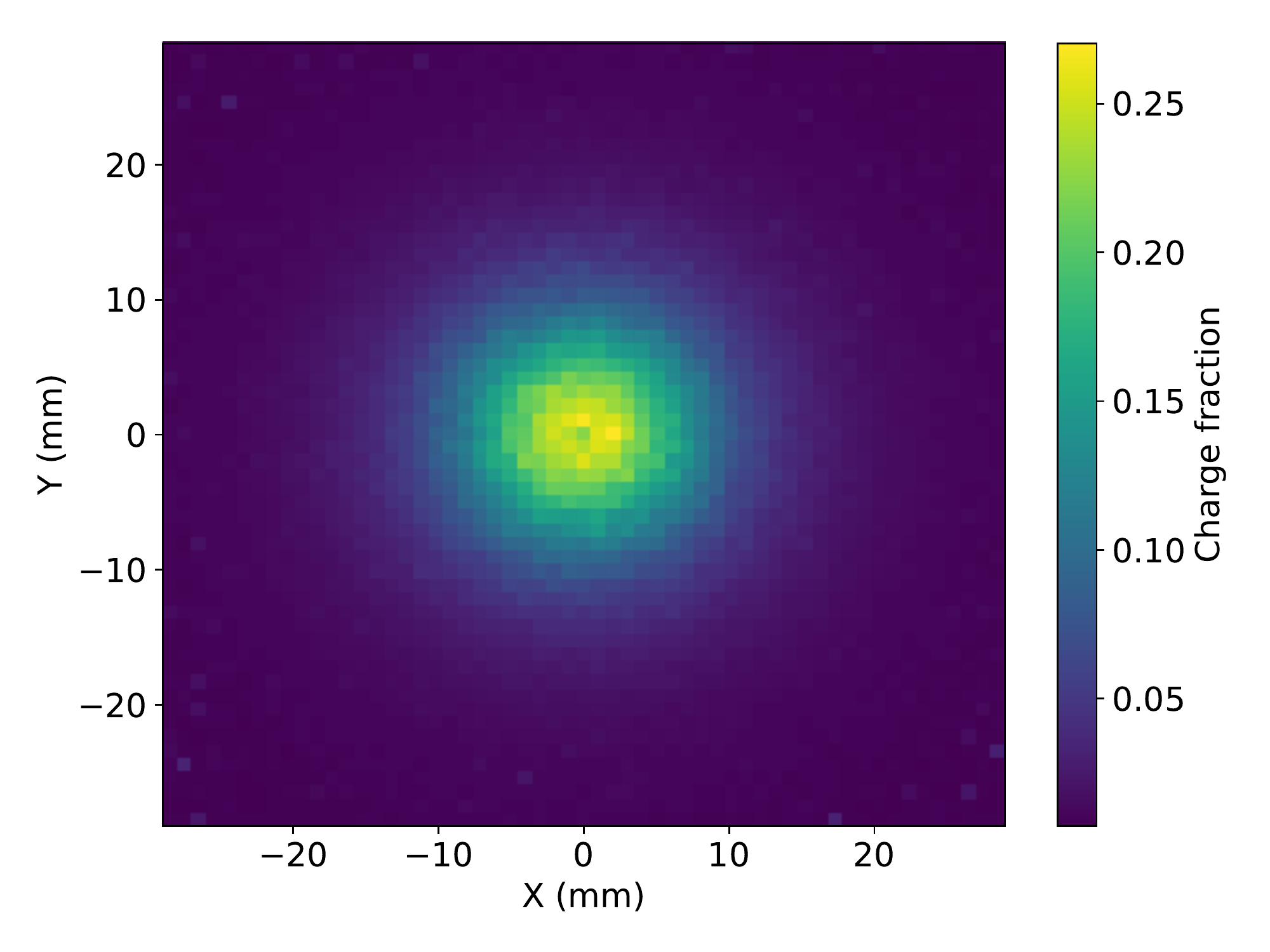}
	\includegraphics[align = c, width=0.4\textwidth]{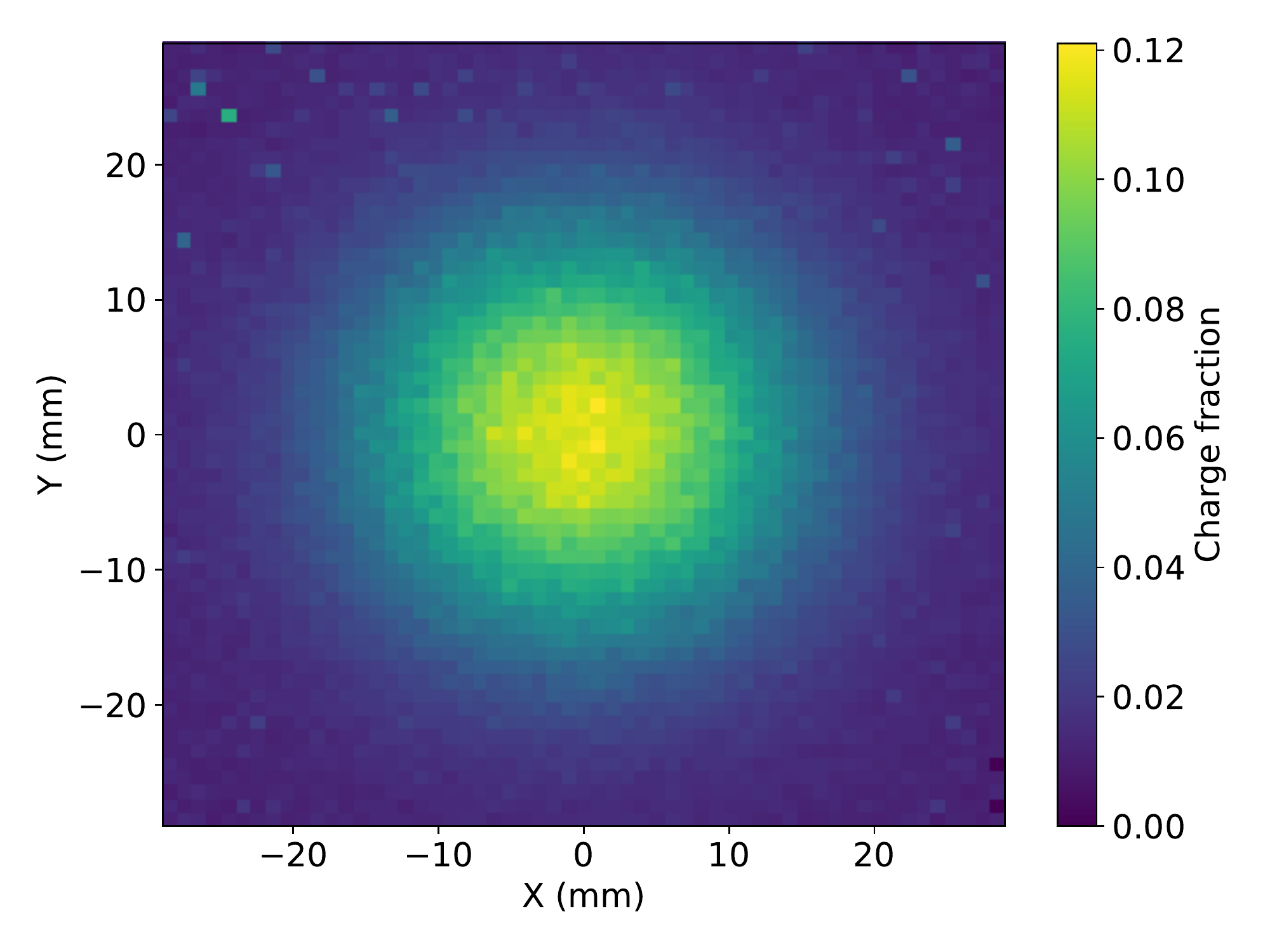}
 \caption{\label{fig:psfDiff} Point spread function (left) for simulated data (top) and standard conditions data (bottom). The charge distribution for a long drift (\SIrange{475}{500}{\micro\second} drift) is shown at the right side. }
  \end{center}
\end{figure}

\begin{figure}[!htb]
  \begin{center}
	\includegraphics[align = c, width=0.4\textwidth]{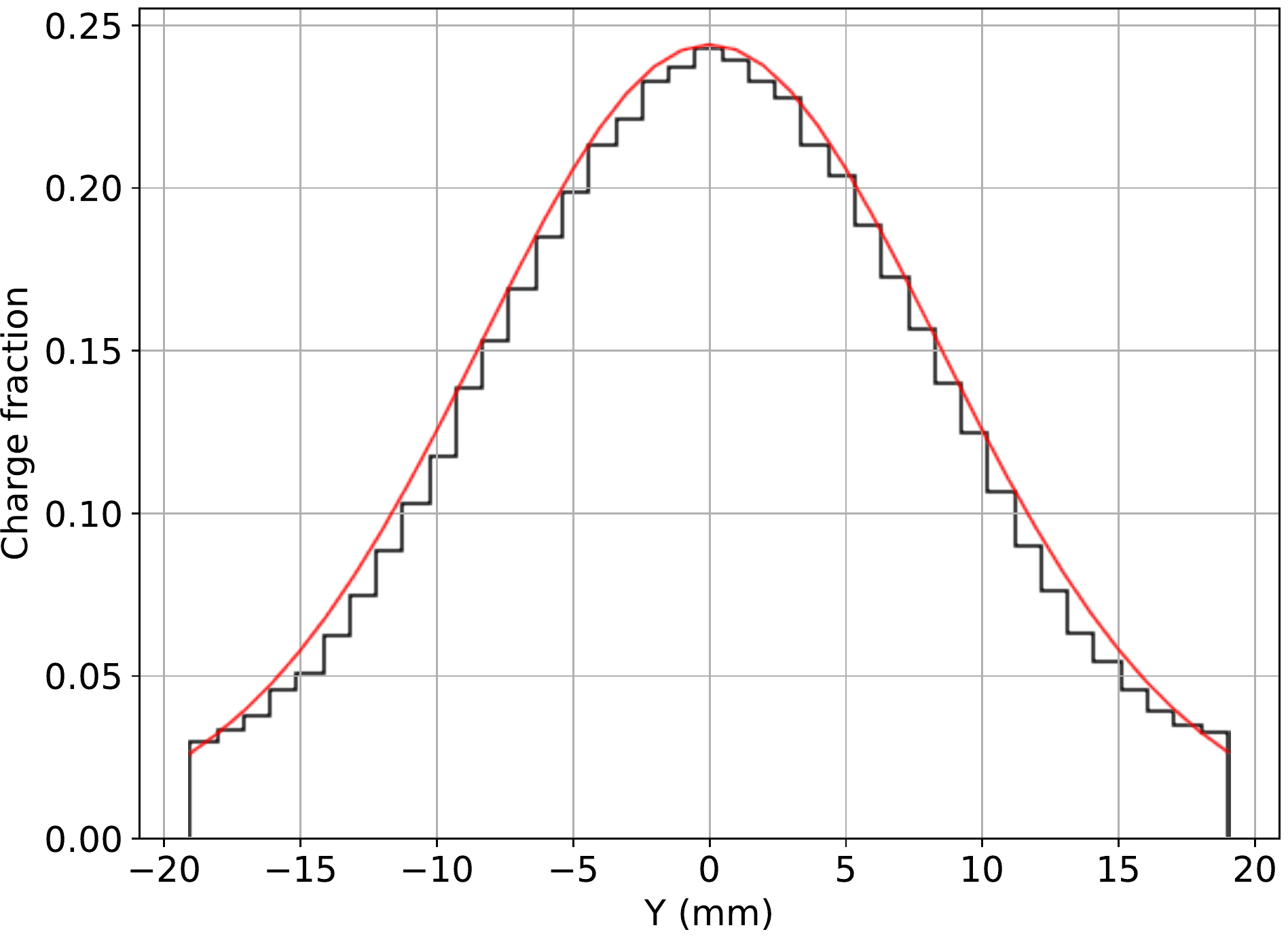}
	\includegraphics[align = c, width=0.4\textwidth]{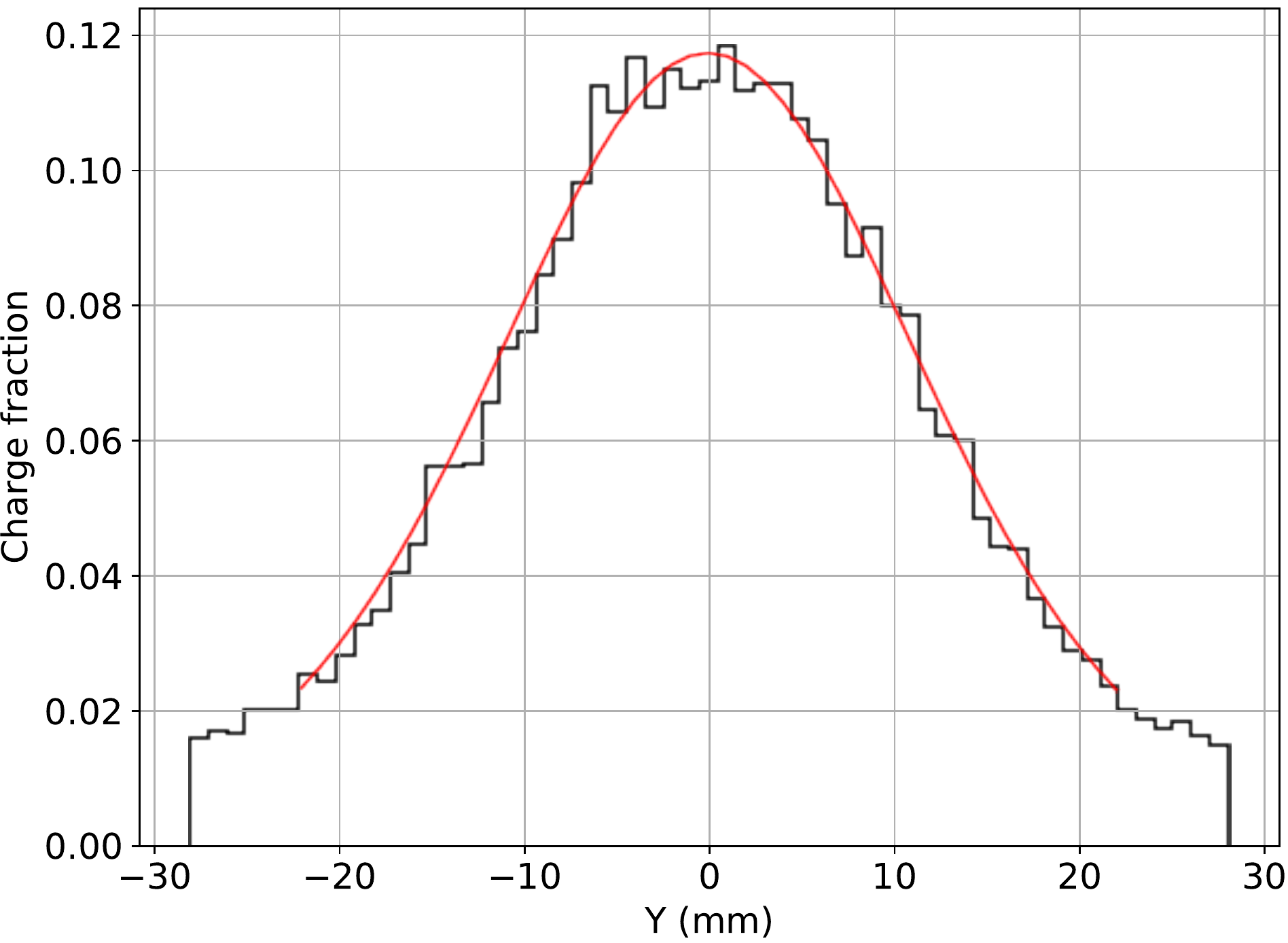}   \caption{\label{fig:psfFit} Transverse slice of the charge distribution at $x$=0 and $z$=[475,500] with the resulting fit (in red). The $\sigma$ obtained from the fit is equal to 6.3 mm for simulated data (left) and 8.2 mm for standard conditions data (right).}
  \end{center}
\end{figure}

For the transverse diffusion measurement, the transverse spread of the drifting electrons has to be computed. The procedure to obtain it is similar to the mean waveform method described in section~\ref{sec:longDiff} to measure the longitudinal diffusion. The data is divided in time intervals and, within each bin, the mean SiPM charge distribution is obtained. 

We define the charge distribution as the \XY\ distribution of each SiPM's position relative to the event's mean position. The relative positions are then weighted by the sensor charge and normalized by the SiPM charge sum. The charge is given by the number of photoelectrons generated by light in each SiPM, integrated over time. The mean position of the event is calculated through a charge-weighted mean of the SiPMs' positions. As before, the mean charge distribution is computed for each drift time interval. 

The charge distribution is a convolution of the point-spread function (PSF) and a gaussian spread due to diffusion. The transverse spread is obtained through a fit to this convolution. The PSF can be estimated from the charge distribution in the short drift region, defined to be the same as the one used for the longitudinal diffusion measurement, namely between 10 and \SI{75}{\micro\second} drift time. The rest of the data is, again, divided in \SI{25}{\micro\second} slices starting at \SI{200}{\micro\second} up to a drift time of \SI{500}{\micro\second}. The light distribution of each drift time bin is fitted to a convolution of the data-extracted PSF with a gaussian. In the fit, an additional consideration was made: only bins with charge higher than 2\% were taken into account. This decision has been made to avoid any effect of diffused light due to light reflections inside the chamber, and to avoid SiPMs whose charge is mostly dark noise. For simulated data, the range of the PSF was only considered up to 20 mm distance. This is because in our simulation the light response of the sensors is only generated for sensors within 20 mm of the electrons. An example of the transverse distributions and fits is shown in figures \ref{fig:psfDiff} and \ref{fig:psfFit} for both data and simulations.

\begin{figure}[!htb]
  \begin{center}
    \includegraphics[height=0.3\textwidth]{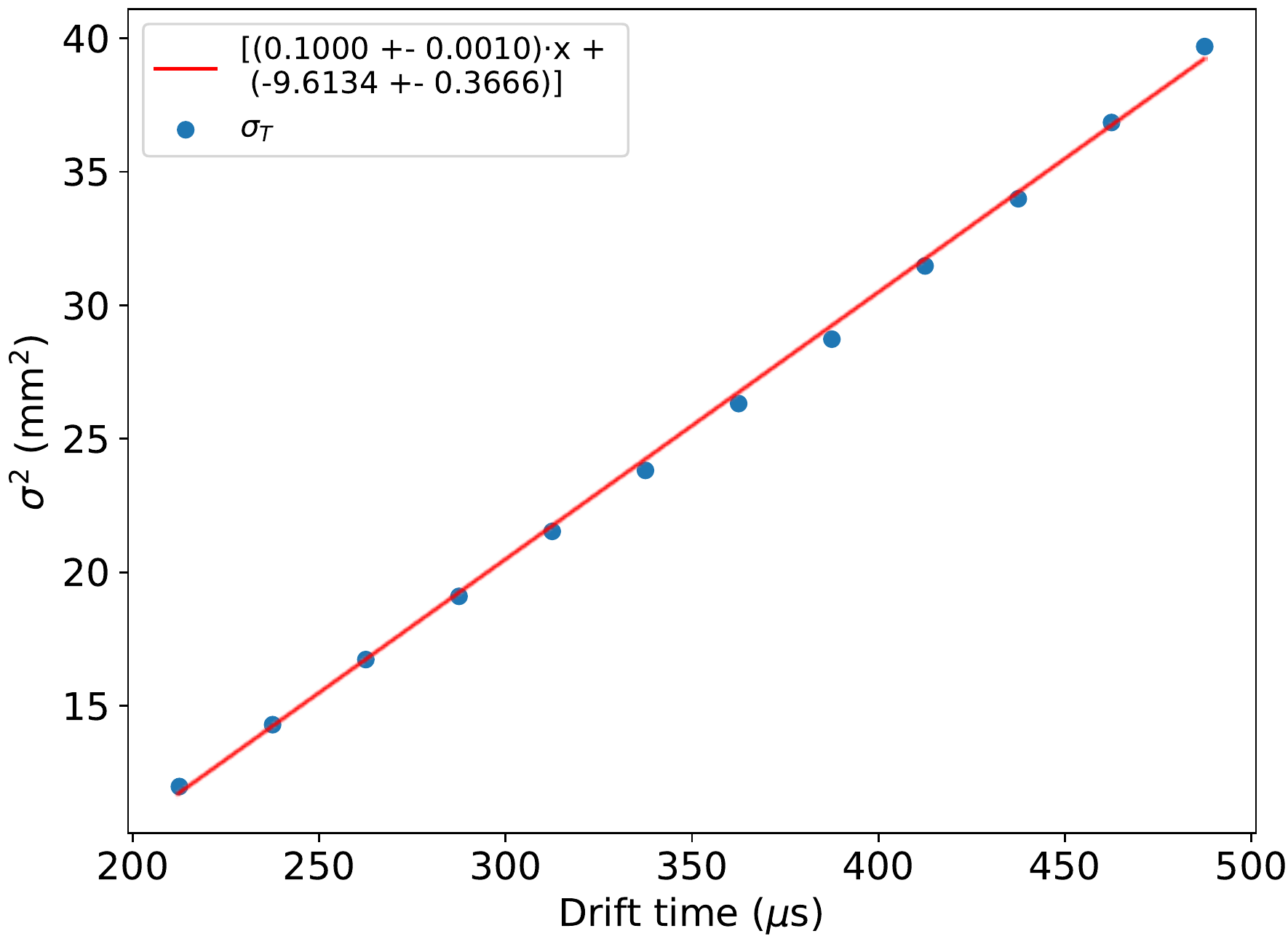}
    \includegraphics[height=0.3\textwidth]{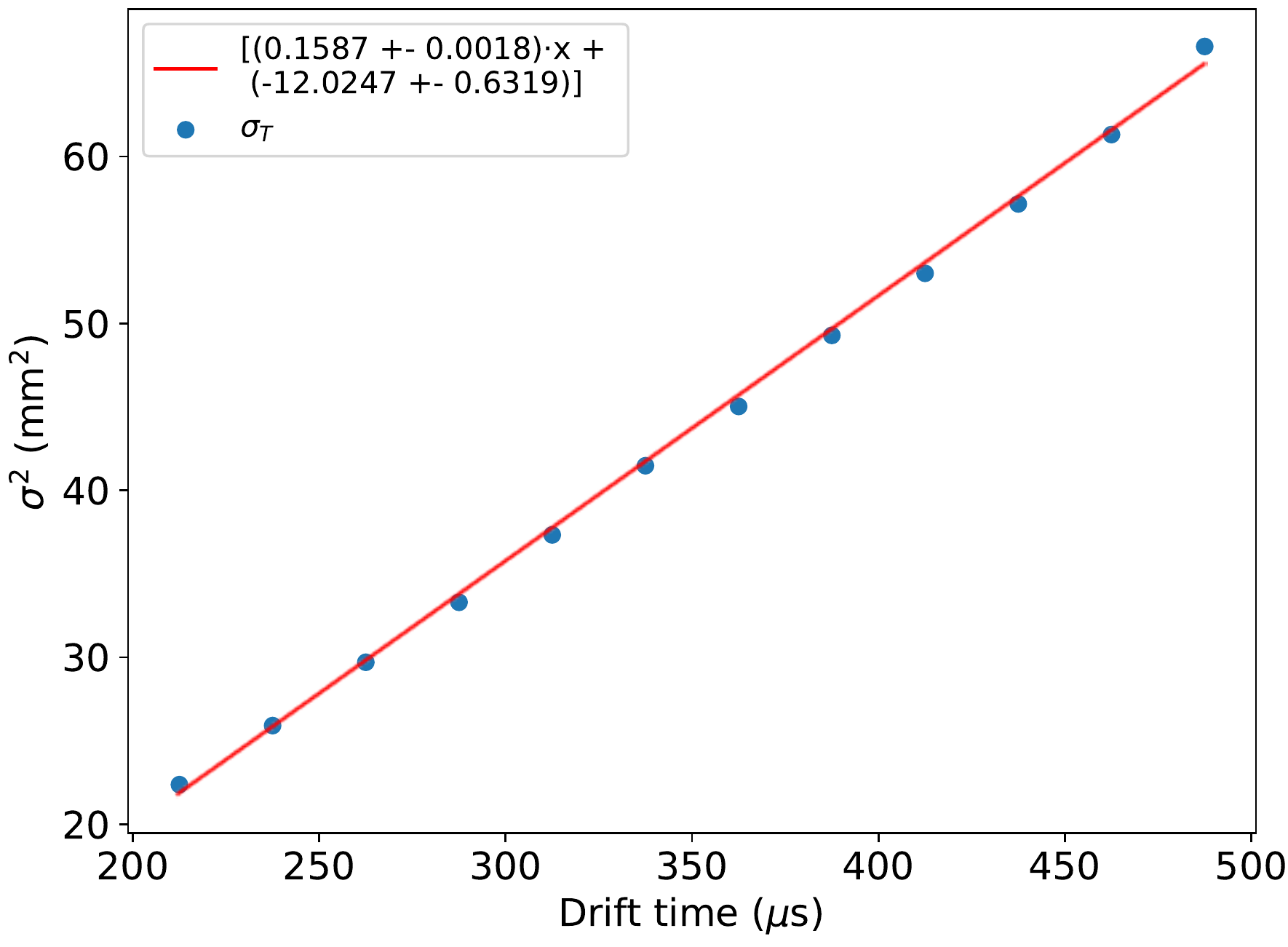}
    \caption{\label{fig:transFit} Transverse spread squared, computed using the mean waveform method, as a function of drift time. The red line shows a linear fit to the data. Left: simulation data. Right: standard conditions data.}
  \end{center}
\end{figure}

\begin{figure}[!htb]
  \begin{center}
    \includegraphics[height=0.5\textwidth]{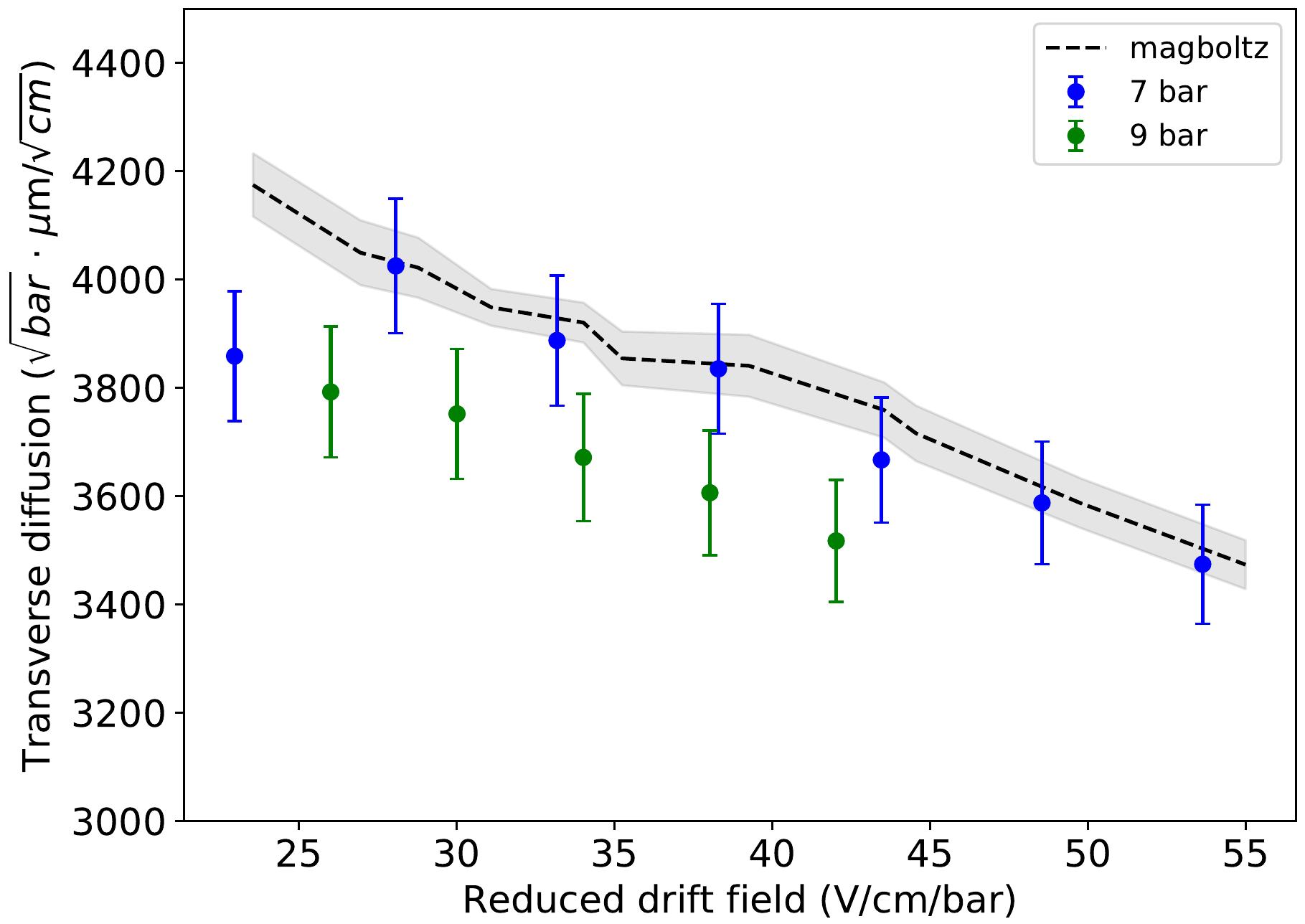}
 \caption{\label{fig:transDiffComparison} Transverse diffusion dependence with the reduced drift field. The results obtained from both 7.2 (blue) and 9.1 (green) bar data are compatible with the Magboltz predictions (black, uncertainties in grey).}
  \end{center}
\end{figure}

As for the longitudinal diffusion, the obtained spreads as a function of the drift time are fitted according to \eq\ \ref{eq:Diffusion}, see \fig\ \ref{fig:transFit}, and the results are transformed following equation~\ref{eq:generalDiffusion}. The results are shown in \fig\ \ref{fig:transDiffComparison}. 

Additionally, the same study of varying the extent of the low-drift window has been repeated also for the transverse diffusion. As for the longitudinal diffusion case, we have conservatively added the maximum difference between the extracted $D_T$~ values as a systematic uncertainty: 0.85\%.

The result obtained was 1000.3 $\pm$ 3.6 (stat.) $\pm$ 10.8 (syst.) \micro m/$\sqrt{\textrm{cm}}$. This result is only 0.01\% away from the result used in the data generation, 1000 \micro m/$\sqrt{\textrm{cm}}$. 

The main source of systematic error on $D_T$~ is given by observed \XY\ differences in the measured charge on the SiPMs depending on the \XY\ position of the events within the detector. These differences are due to several detector imperfections such as irregularities in the EL mesh or TPB inhomogeneities. The impact of these differences on the diffusion measurements has been evaluated by dividing the detector in four \XY\ quadrants, and applying the $D_T$~ extraction method separately for each quadrant. The maximum difference in the $D_T$~ measurements from the four quadrants is taken as a systematic error. For the standard conditions data, this procedure resulted in a 1.8\% systematic error.

Although for the longitudinal diffusion measurement the mean waveform method was cross-checked with an event-by-event approach, a similar check is not possible for the transverse diffusion case. The reason is that the separation between SiPMs, 10 mm apart, dominates any intent of calculating the RMS on an event-by-event basis.

A comparison of our $D_T$~ results with Magboltz simulations is shown on \fig\ \ref{fig:transDiffComparison}. As can be seen, the results are compatible with the Magboltz predictions with a mean difference of only 3.1\%, being 7.6\% in the worst case and 0.01\% in the best one. Again, the fact that our results and Magboltz predictions at 1 bar are compatible with each other confirms the validity of P-scaling. Our data hints towards \NewSevenBarPressureRunII\ $D_T$~ results being somewhat higher than the \NewNineBarPressureRunII\ bar $D_T$~ ones. The cause of the discrepancy remains to be studied in detail although it may be related to the  lifetime dependence with the transverse position that has been observed in the \NewSevenBarPressureRunII\ data but not in the \NewNineBarPressureRunII\ one, as stated in \cite{Martinez-Lema:2018ibw}. The irregular lifetime may affect the charge distribution of the SiPMs having an effect in the outcome of the sensor's charge cut, which was fixed for the all the datasets. In this work lifetime has been considered uniform for simplicity and the impact of its inhomogeneous distribution should be reflected in the systematic error obtained through the four quadrant evaluation. Nevertheless, results for the studied pressures are completely compatible within the considered errors.


\section{Conclusions}
\label{sec.conclu}
A high precision characterization of different electron transport properties for xenon gas at high pressure has been presented in this work. Concretely, drift velocity, longitudinal diffusion and transverse diffusion have all been measured simultaneously, for the first time, using the same experimental setup. The analysis uses \Kr{83m} calibration data taken in the NEXT-White (NEW) detector at the Laboratorio Subterr\'aneo de Canfranc (LSC) between September and October 2017, during Run II of the detector.

A study of the transport parameters dependence with reduced drift field has been done for two different gas pressures, \NewSevenBarPressureRunII\ and \NewNineBarPressureRunII. The results, after applying density scalings, are compatible, within 5\% in most cases, with Magboltz simulations of pure xenon carried out at 1 bar and in the same reduced drift field conditions. In the drift velocity case, the average deviation between Magboltz and our results is of 1.2\%, with a maximum deviation of 1.8\%. The measured longitudinal diffusion shows an average deviation of 3.5\%, while the maximum deviation is 6.6\%. Finally, the mean deviation observed in the transverse diffusion is of 3.1\%, with a maximum of 7.6\%.

\begin{figure}[!htb]
  \begin{center}
   \hspace{0.5em}
	\includegraphics[align = c, width=0.536\textwidth]{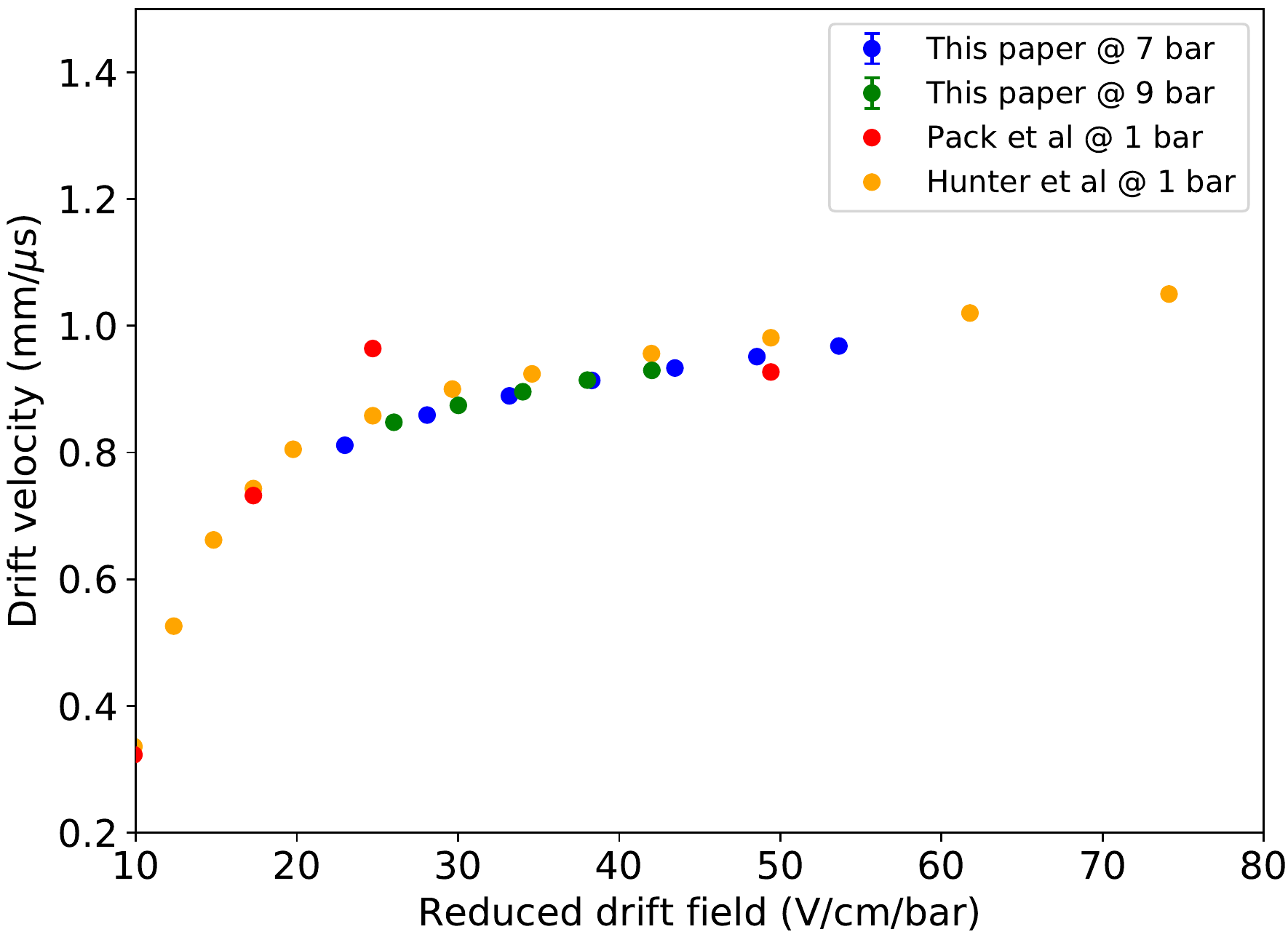}\vspace{1em}
	\includegraphics[align = c, width=0.55\textwidth]{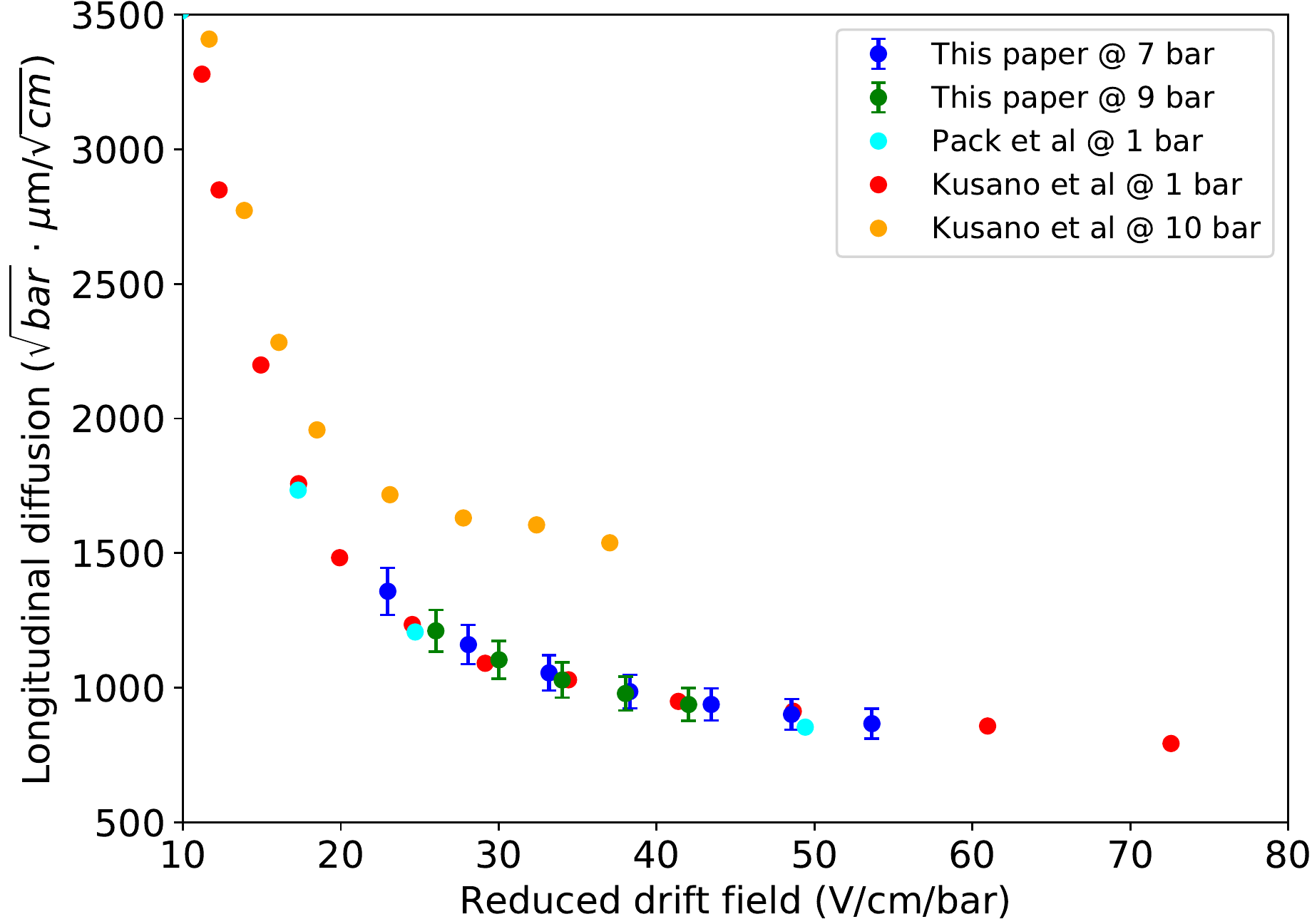}\vspace{1em}
	\includegraphics[align = c, width=0.55\textwidth]{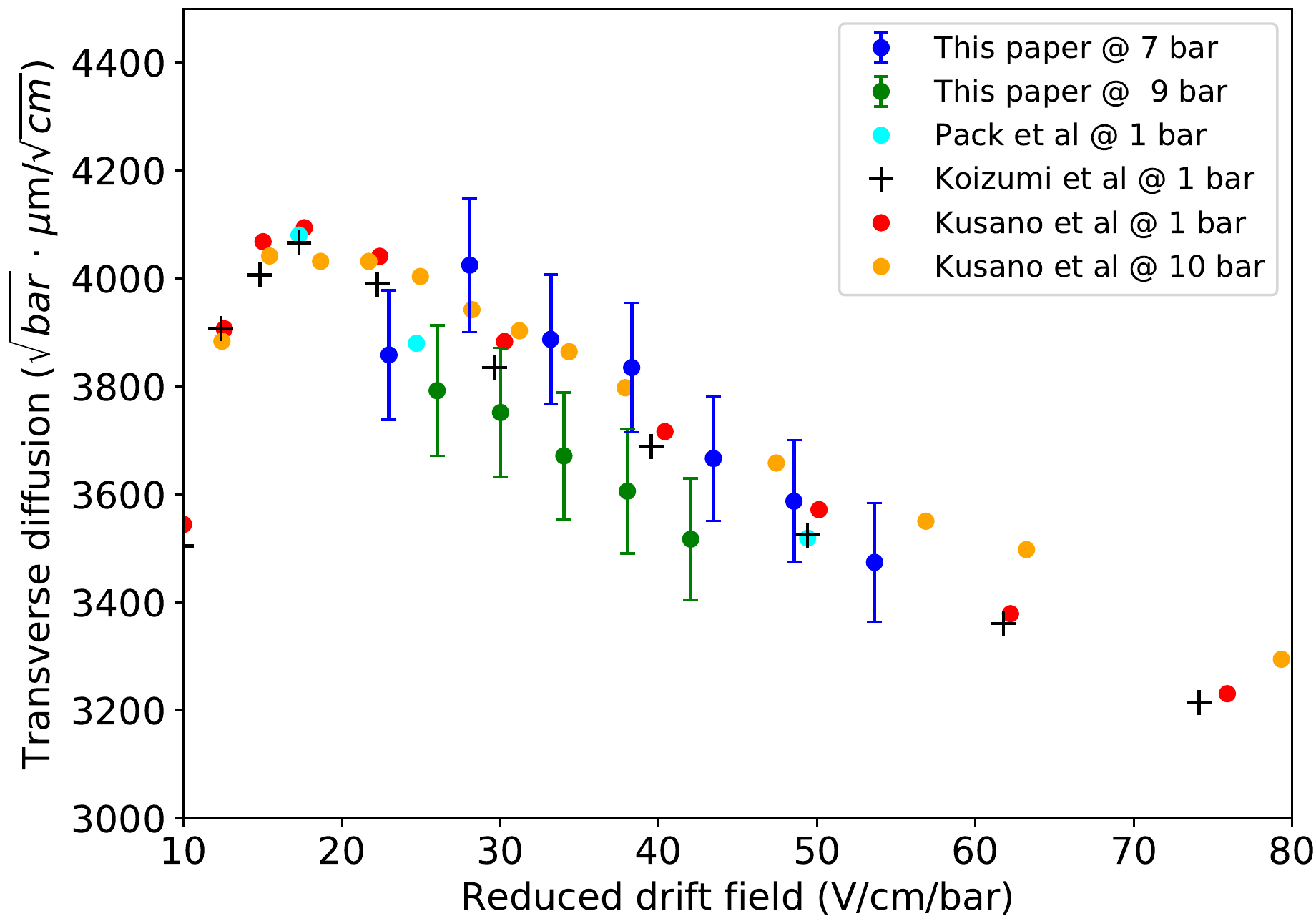}   \caption{\label{fig:comparison} Comparison between experimental measurements at different pressures of drift velocity (top), longitudinal diffusion (middle) and transverse diffusion (bottom).}
  \end{center}
\end{figure}

As can be seen in \fig\ \ref{fig:comparison}, the results are compatible with previous experimental measurements carried out at different pressures \cite{Pack:1962, Pack:1992, Hunter:1988, Koizumi:1986, Kusano:2013, Bowe:1960, Kobayashi:2006, Patrick:1991, English:1953}. This level of agreement shows that pressure scaling of the diffusion parameters can be trusted for these operating pressures, suggesting that the role of dimers and higher order xenon clusters in electron transport is still minor at pressures up to $\sim$10 bar, at least. Therefore, we are unable to reproduce the results obtained in \cite{Kusano:2013}, where the reduced longitudinal diffusion coefficients showed a dependence with pressure when working in the same reduce field range studied here.

With regard to the implications of this result for the NEXT experiment, the measured longitudinal diffusion applicable to most of the 2017 \NEW\ calibration campaign is \StandardLDRunII, while the transverse diffusion value is \StandardTDRunII. The NEXT collaboration will work at a minimum of \NewMinimumPressure\ for the upcoming physics runs. This pressure implies a longitudinal diffusion of \LDTenBarRunII\ and a transverse diffusion of \TDTenBarRunII. 
These values are compatible with the expectations and the design of the NEXT-100 detector \cite{Alvarez:2011my, Alvarez:2012haa, Gomez-Cadenas:2013lta}.

\acknowledgments
The NEXT Collaboration acknowledges support from the following agencies and institutions: the European Research Council (ERC) under the Advanced Grant 339787-NEXT; the European Union's Framework Programme for Research and Innovation Horizon 2020 (2014-2020) under the Marie Sk\l{}odowska-Curie Grant Agreements No. 674896, 690575 and 740055; the Ministerio de Econom\'ia y Competitividad of Spain under grants FIS2014-53371-C04, the Severo Ochoa Program SEV-2014-0398 and the Mar\'ia de Maetzu Program MDM-2016-0692; the GVA of Spain under grants PROMETEO/2016/120 and SEJI/2017/011; the Portuguese FCT and FEDER through the program COMPETE, projects PTDC/FIS-NUC/2525/2014 and UID/FIS/04559/2013; the U.S.\ Department of Energy under contracts number DE-AC02-07CH11359 (Fermi National Accelerator Laboratory), DE-FG02-13ER42020 (Texas A\&M) and de-sc0017721 (University of Texas at Arlington); and the University of Texas at Arlington. We also warmly acknowledge the Laboratorio Nazionale di Gran Sasso (LNGS) and the Dark Side collaboration for their help with TPB coating of various parts of the NEXT-White TPC. Finally, we are grateful to the Laboratorio Subterr\'aneo de Canfranc for hosting and supporting the NEXT experiment.

\clearpage

\bibliography{krDiff}
\end{document}